\documentclass[twocolumn, prb]{revtex4}
\usepackage{amsfonts}
\usepackage{amssymb}
\usepackage{amsmath}
\usepackage{color}
\usepackage{graphicx}

\begin{document}

\title{On the effective shear speed in 2D phononic crystals}

\author{A.A. Kutsenko$^{a}$, A.L. Shuvalov$^{a,*}$, A.N. Norris$^{a,b}$, O. Poncelet$^{a}$}
\affiliation{ $^{a}\ $Universit\'{e} de Bordeaux, Institut de
M\'{e}canique et d'Ing\'{e}nierie de Bordeaux,
UMR 5295, Talence 33405, France\\
$^{b}$ Mechanical and Aerospace Engineering, Rutgers University,
Piscataway, NJ 08854-8058, USA\\
$^{*}$Corresponding author, email: a.shuvalov@lmp.u-bordeaux1.fr}

\begin{abstract}
The quasistatic limit of the antiplane shear-wave speed ('effective speed') $%
c$ in 2D periodic lattices is studied. Two new closed-form estimates
of $c$ are derived by employing two different analytical approaches.
The first proceeds from a standard background of the plane wave
expansion (PWE). The second is a new approach, which resides in
$\mathbf{x}$-space and centers on the monodromy matrix (MM)
introduced in the 2D case as the multiplicative integral, taken in
one coordinate, of a matrix with components being the operators with
respect to the other coordinate. On the numerical side, an efficient
PWE-based scheme for computing $c$ is proposed and implemented. The
analytical and numerical findings are applied to several examples of
2D square lattices with two and three high-contrast components, for
which the new PWE and MM estimates are compared with the numerical
data and with some known approximations. It is demonstrated that the
PWE estimate is most efficient in the case of densely packed stiff
inclusions, especially when they form a symmetric lattice, while in
general it is the MM estimate that provides the best overall fitting
accuracy.\medskip
\end{abstract}

\maketitle

\section{Introduction}

\label{sec1}

Effective material properties of composites have been and remain a topic of
much interest in micromechanics, see the reviews \cite{BM,MMP,PA}. The
recent surge of research into the properties of metamaterials and phononic
crystals has heightened attention, particularly for periodic systems. In
this context, considerable work has been done on the low-frequency, or
quasistatic, limit of the antiplane shear-wave speed in 2D periodic
structures (referred to as the 'effective speed' $c$ in the following; note
that this value also yields the limit of the fundamental velocity branch of
shear plate waves). A natural tool for tackling the problems with
periodicity is the plane-wave expansion (PWE). An explicit PWE-expression of
the effective speed $c$ via an infinite sum of Fourier coefficients has been
obtained in \cite{KAG} and was broadly used afterwards for computing $c$ in
various periodic materials. Note some other semi-analytical techniques that
were used for numerical evaluation of $c,$ such as scaling \cite{ABDW} and
mixed-variational \cite{NWSA} methods. In turn, the multiple-scattering
theory (MST), which deals directly with the inclusion/matrix boundary
problem, has proved to be expedient for deriving the effective speed in an
approximate but closed form. By means of MST, such simple (in appearance)
estimate of $c$, which had been known for certain statistically uniform
models in micromechanics, was recently extended to phononic crystals with a
periodic microstructure of inclusions \cite{MLWS,TS,SMLW,TS1}.

The main results of the present paper are concerned with both analytical and
numerical aspects of the problem of evaluating $c$. The analytical
development aims at finding approximations of $c$ by means independent of
the MST. The starting point is the general expression for $c$ in the
operator form that may be further specialized to either Fourier space or $%
\mathbf{x}$-space. On this basis, we provide two new closed-form estimates
of $c$ derived by employing two different analytical frameworks. The first
is the PWE approach, which commences from the formula of \cite{KAG}. The
second is a completely new approach based on the monodromy matrix (MM),
which is a fundamental object for the 1D periodic problems (cf. the
state-vector formalism) but it has not seen much, if any, far-reaching
application in 2D. Here the MM is introduced as a multiplicative integral,
taken with respect to one coordinate, of the matrix with components defined
as operators acting on the functions of the other coordinate. On the
numerical side, we develop an efficient PWE-based scheme for computing $c$,
in which the matrix inversion is replaced by the power series that is
judiciously gauged for its faster convergence. The results are applied to
several examples of 2D square lattices consisting of two and three
high-contrast components with filling fractions $f$, for which the new PWE
and MM estimates of $c\left( f\right) $ and its known MST estimate are
compared against the benchmark of the numerically computed $c\left( f\right)
.$ In brief, it is demonstrated that the PWE estimate is efficient in the
case of densely packed stiff inclusions (where the MST estimate fails) and
is particularly useful for the symmetric binary lattices invariant to
interchanging their components (in which case the MST formula is ambiguous);
but it is the MM estimate that provides the best overall fit over various
lattice configurations considered.

The paper is organized as follows. The background expression for $c$
is presented in \S \ref{sec2}. The PWE and MM closed-form estimates
of $c$ are derived in \S \ref{sec3}. The numerical scheme used for
computing $c$ is described in \S \ref{sec4}. Application of
analytical and numerical results to 2D square lattices is discussed
in \S \ref{sec5}. Concluding remarks are presented in \S \ref{sec6}.
Appendix expands on the convergence of the implemented numerical
scheme.

\section{Background: exact expression for effective speed}

\label{sec2}

Consider a 2D periodic locally isotropic medium with the density $\rho
\left( \mathbf{x}\right) $ and the shear coefficient $\mu (\mathbf{x}),$
which are real positive piecewise continuous functions satisfying
\begin{eqnarray}
\rho \big(\mathbf{x+}\sum\nolimits_{j=1}^{2}n_{j}\mathbf{a}_{j}\big) &=&\rho
(\mathbf{x}),  \notag \\
\mu \big(\mathbf{x+}\sum\nolimits_{j=1}^{2}n_{j}\mathbf{a}_{j}\big) &=&\mu (%
\mathbf{x})  \label{1}
\end{eqnarray}%
for any $\mathbf{x}\in
\mathbb{R}
^{2},$ $n_{j}\in
\mathbb{Z}
$ and some linear independent translation vectors $\mathbf{a}_{j}\in
\mathbb{R}
^{2}$ that form the irreducible unit cell $\mathbf{T}=\sum%
\nolimits_{j=1}^{2}t_{j}\mathbf{a}_{j}$ ($t_{j}\in \left[ 0,1\right] $) of
the 2D periodic lattice. Let $\mathbf{e}_{1},\mathbf{e}_{2}\equiv \left\{
\mathbf{e}_{j}\right\} $ be an orthonormal base in $%
\mathbb{R}
^{2},$ and $\mathbf{x\cdot y}=\sum_{i=1}^{2}x_{i}y_{i}$ be the scalar
product in $%
\mathbb{R}
^{2},$ where $x_{i}$ are the coordinates of an arbitrary vector $\mathbf{x}$
with respect to $\left\{ \mathbf{e}_{j}\right\} $. Denote%
\begin{equation}
\mathbf{a}_{j}=\mathbf{Ae}_{j},\ \mathbf{b}_{j}=\left( \mathbf{A}%
^{-1}\right) ^{\mathrm{T}}\mathbf{e}_{j},\ \mathbf{g}=\sum%
\nolimits_{j=1}^{2}2\pi n_{j}\mathbf{b}_{j},  \label{2}
\end{equation}%
where $\mathbf{a}_{j}\cdot \mathbf{b}_{k}=\delta _{jk}$ $\left(
j,k=1,2\right) ,\ $ $^{\mathrm{T}}$ means transpose, and $\mathbf{g}%
=\sum\nolimits_{j=1}^{2}g_{j}\mathbf{e}_{j}$ is the reciprocal lattice
vector whose components $\left( g_{1},g_{2}\right) $ in $\left\{ \mathbf{e}%
_{j}\right\} $ take all values from the set $\Gamma =2\pi \left( \mathbf{A}%
^{-1}\right) ^{\mathrm{T}}%
\mathbb{Z}
^{2}$. In the following, the Fourier coefficients of a periodic function $f(%
\mathbf{x})$ are indicated by a hat:%
\begin{gather}
f(\mathbf{x})=\sum\nolimits_{\mathbf{g}}\widehat{f}(\mathbf{g})e^{i\mathbf{%
g\cdot x}}\ \Leftrightarrow  \notag \\
\ \widehat{f}\left( \mathbf{g}\right) =\frac{1}{\left\vert \mathbf{T}%
\right\vert }\int_{\mathbf{T}}f\left( \mathbf{x}\right) e^{-i\mathbf{g\cdot x%
}}\mathrm{d}\mathbf{x}\equiv \left\langle f(\mathbf{x})e^{-i\mathbf{g\cdot x}%
}\right\rangle ;  \label{4}
\end{gather}%
and the same notation $\left( \cdot ,\cdot \right) $ is used for the scalar
products in $\mathbf{g}$-space and in $L^{2}\left( \mathbf{T}\right) $:
\begin{multline}
\left( f,h\right) =\sum\nolimits_{\mathbf{g}}\widehat{f}(\mathbf{g})\widehat{%
h}(-\mathbf{g})\\ =\frac{1}{\left\vert \mathbf{T}\right\vert }\int_{\mathbf{T}%
}f\left( \mathbf{x}\right) h^{\ast }\left( \mathbf{x}\right) \mathrm{d}%
\mathbf{x}\equiv \langle fh^{\ast }\rangle .  \label{2.1}
\end{multline}

The antiplane\emph{\ }time-harmonic displacement $v\left( \mathbf{x}%
,t\right) =v\left( \mathbf{x}\right) e^{\mathbf{-}i\omega t}$ is determined
by the wave equation%
\begin{equation}
\mathbf{\nabla }\cdot \left( \mu (\mathbf{x})\mathbf{\nabla }v(\mathbf{x}%
)\right) =-\rho (\mathbf{x})\omega ^{2}v(\mathbf{x})  \label{3}
\end{equation}%
with periodic $\rho \left( \mathbf{x}\right) $ and $\mu (\mathbf{x})$. By
this periodicity, $v(\mathbf{x})=u(\mathbf{x})e^{i\mathbf{k\cdot x}}$ where $%
u\left( \mathbf{x}\right) $ is periodic and $\mathbf{k}=k\mathbf{\kappa }$ ($%
\left\vert \mathbf{\kappa }\right\vert =1$) is the Floquet vector, and so
Eq. (\ref{3}) can be cast as%
\begin{gather}
({\mathcal{C}}_{0}+{\mathcal{C}}_{1}+{\mathcal{C}}_{2})u=\rho \omega ^{2}u\
\ \ \mathrm{with\ \ }{\mathcal{C}}_{0}u=-\mathbf{\nabla }(\mu \mathbf{\nabla
}u),  \notag \\
\ \ {\mathcal{C}}_{1}u=-i\mathbf{k}\cdot (\mu \mathbf{\nabla }u+\mathbf{%
\nabla }(\mu u)),\ \ {\mathcal{C}}_{2}u=k^{2}\mu u.  \label{501}
\end{gather}%
To find the effective speed $c(\mathbf{\kappa })=\lim_{\omega ,k\rightarrow
0}\omega \left( \mathbf{k}\right) /k,$ consider the asymptotics $\omega
^{2}=\omega _{0}^{2}+\omega _{1}^{2}+\omega _{2}^{2}+O(k^{3}).$ It is
evident that $\omega _{0}^{2}=0$ is an eigenvalue of (\ref{501}) with the
eigenvector $u_{0}=1.$ Therefore perturbation theory yields%
\begin{equation}
\omega _{1}^{2}=\frac{\left( {\mathcal{C}}_{1}u_{0},u_{0}\right) }{\left(
\rho u_{0},u_{0}\right) },\ \omega _{2}^{2}=\frac{\left( {\mathcal{C}}%
_{2}u_{0},u_{0}\right) -\left( {\mathcal{C}}_{0}^{-1}{\mathcal{C}}_{1}u_{0},{%
\mathcal{C}}_{1}u_{0}\right) }{\left( \rho u_{0},u_{0}\right) },  \label{502}
\end{equation}%
where $\left( \rho u_{0},u_{0}\right) =\langle \rho \rangle $ and$\ \left( {%
\mathcal{C}}_{2}u_{0},u_{0}\right) =k^{2}\langle \mu \rangle .$ Note that $%
\left( {\mathcal{C}}_{1}u_{0},u_{0}\right) =0$ by periodicity of $\mu $,
hence $\omega _{1}^{2}=0$ and the operator ${\mathcal{C}}_{0}^{-1}$ is
defined in the subspace $L_{0}^{2}$ of all functions $f$ orthogonal to 1,
i.e., such that $\left\langle f\right\rangle =0.$ Thus $c^{2}(\mathbf{\kappa
})$ is expressed via the averaged density $\langle \rho \rangle ~\left( =%
\widehat{\rho }\left( 0\right) \right) $ and the effective shear coefficient
$\mu _{\mathrm{eff}}(\mathbf{\kappa })$ as follows:%
\begin{gather}
c^{2}(\mathbf{\kappa })=\frac{\mu _{\mathrm{eff}}\left( \mathbf{\kappa }%
\right) }{\langle \rho \rangle },\ \mu _{\mathrm{eff}}(\mathbf{\kappa }%
)=\left\langle \mu \right\rangle -M(\mathbf{\kappa })  \notag \\
\mathrm{with\ }M\left( \mathbf{\kappa }\right)
=\sum\nolimits_{i,j=1}^{2}\left( {\mathcal{C}}_{0}^{-1}\frac{\partial \mu }{%
\partial x_{i}},\frac{\partial \mu }{\partial x_{j}}\right) \kappa
_{i}\kappa _{j}.  \label{6}
\end{gather}
The operator ${\mathcal{C}}_{0}^{-1}$ is compact and also self-adjoint and
positive, whence
\begin{equation}
c^{2}(\mathbf{\kappa })\leq \left\langle \mu \right\rangle /\left\langle
\rho \right\rangle .  \label{10}
\end{equation}

It is worth emphasizing that the perturbation theory enables an efficient
shortcut to an explicit expression of the effective speed $c$, in which the
quadratic form $M\left( \mathbf{\kappa }\right) $ may be specialized to
either $\mathbf{g}$- or $\mathbf{x}$-space. Taking a double Fourier
expansion of (\ref{6}$_{2}$),%
\begin{eqnarray}
M(\mathbf{\kappa }) &=&\sum\nolimits_{\mathbf{g,g}^{\prime }\in \Gamma
\backslash \left\{ \mathbf{0}\right\} }\widehat{\mu }\left( \mathbf{g}%
\right) \widehat{\mu }\left( -\mathbf{g}^{\prime }\right) \left( \mathbf{g}%
\cdot \mathbf{\kappa }\right) \left( \mathbf{g}^{\prime }\cdot \mathbf{%
\kappa }\right)  \notag \\
&&\times \left( \widehat{\mu }\left( \mathbf{g-g}^{\prime }\right) \mathbf{%
g\cdot \mathbf{g}^{\prime }}\right) ^{-1}  \label{7}
\end{eqnarray}
provides the PWE-representation of $c^{2}(\mathbf{\kappa })$ obtained in
\cite{KAG}. Viewing Eq. (\ref{6}) along with the equation ${\mathcal{C}}%
_{0}h=\partial \mu /\partial x_{i}$ in $\mathbf{x}$-space is precisely
equivalent to the formulation of quasistatic limit by the scaling approach,
see \cite{BP}. The above derivation, taking a few lines, does not need the
scaling ansatz. Moreover, while the central point of the scaling approach is
the use of the Fredholm alternative (Lemma 1 in Ch. 4 of \cite{BP}), the
same is inherent to the perturbation theory 'by construction' whereby the
eigenfunction perturbations are confined via the operator $C_{0}^{-1}$ to
the subspace $L_{0}^{2}$ orthogonal to the unperturbed eigenfunction (in $%
\mathbf{g}$-space, this is implied by the summation over
$\mathbf{g}\in \Gamma \backslash \left\{ \mathbf{0}\right\} $ in Eq.
(\ref{7})). Finally, note that Eq. (\ref{6}) can be further
developed by using the monodromy-matrix approach, see \S
\ref{sec3.2}.

\section{Estimates of the effective speed}

\label{sec3}

\subsection{PWE estimate}

\label{sec3.1}

Eq. (\ref{7}) of \cite{KAG} defines $M\left( \mathbf{\kappa }\right) $ as a
scalar product in the Fourier space $l^{2}\left( \Gamma \backslash \left\{
\mathbf{0}\right\} \right) $,
\begin{equation}
M(\mathbf{\kappa })=\left( \mathbf{B}^{-1}\mathbf{d,d}\right) ,  \label{8}
\end{equation}%
where $\mathbf{B}$ is an infinite matrix and $\mathbf{d}$ an infinite vector
with components%
\begin{gather}
\mathbf{B}\equiv \left( B\left[ \mathbf{g,g}^{\prime }\right] \right) _{%
\mathbf{g,g}^{\prime }\in \Gamma \backslash \left\{ \mathbf{0}\right\} }:\ B%
\left[ \mathbf{g,g}^{\prime }\right] =\widehat{\mu }\left( \mathbf{g-g}%
^{\prime }\right) \mathbf{g\cdot g}^{\prime };  \notag \\
\mathbf{d}\equiv \left( d\left( \mathbf{g}\right) \right) :\ d\left( \mathbf{%
g}\right) =\widehat{\mu }\left( \mathbf{g}\right) \mathbf{g}\cdot \mathbf{%
\kappa .}  \label{9}
\end{gather}%
By definition (\ref{9}$_{1}$), $\mathbf{B}^{-1}$ is a compact operator in $%
l^{2}\left( \Gamma \backslash \left\{ \mathbf{0}\right\} \right) .$ Let us
further cast $\mu (\mathbf{x})$ in the form
\begin{equation}
\mu (\mathbf{x})=\mu _{0}+\mu _{\Delta }(\mathbf{x}),  \label{11}
\end{equation}%
where $\mu _{0}$ is some positive constant and hence $\widehat{\mu }\left(
\mathbf{g-g}^{\prime }\right) =\mu _{0}\delta _{\mathbf{gg}^{\prime }}+%
\widehat{\mu }_{\Delta }\left( \mathbf{g-g}^{\prime }\right) $. Denote%
\begin{gather}
\mathbf{C}(\mu _{0})\equiv \left( C\left[ \mathbf{g,g}^{\prime }\right]
\right) _{\mathbf{g,g}^{\prime }\in \Gamma \backslash \left\{ \mathbf{0}%
\right\} }\mathrm{:}\ ~  \notag \\
C\left[ \mathbf{g,g}^{\prime }\right] =\frac{\widehat{\mu }_{\Delta }}{\mu
_{0}}\left( \mathbf{g-g}^{\prime }\right) \frac{\mathbf{g}}{\left\vert
\mathbf{g}\right\vert }\cdot \frac{\mathbf{g}^{\prime }}{\left\vert \mathbf{g%
}^{\prime }\right\vert };  \notag \\
\mathbf{D}\equiv \mathrm{diag}\left( \left\vert \mathbf{g}\right\vert
\right) _{\mathbf{g}\in \Gamma \backslash \left\{ \mathbf{0}\right\} };
\notag \\
\mathbf{f=D}^{-1}\mathbf{d}\equiv \left( \widehat{f}\left( \mathbf{g}\right)
\right) _{\mathbf{g}\in \Gamma \backslash \left\{ \mathbf{0}\right\} }%
\mathrm{:~}\ \   \notag \\
\widehat{f}\left( \mathbf{g}\right) =\widehat{\mu }\left( \mathbf{g}\right)
\frac{\mathbf{g}}{\left\vert \mathbf{g}\right\vert }\cdot \mathbf{\kappa }=%
\widehat{\mu }_{\Delta }\left( \mathbf{g}\right) \frac{\mathbf{g}}{%
\left\vert \mathbf{g}\right\vert }\cdot \mathbf{\kappa ;}  \notag \\
\left( \mathbf{f,f}\right) =\sum\nolimits_{\mathbf{g}\neq \mathbf{0}%
}\left\vert \widehat{\mu }\left( \mathbf{g}\right) \right\vert ^{2}\frac{%
\left( \mathbf{g}\cdot \mathbf{\kappa }\right) ^{2}}{\left\vert \mathbf{g}%
\right\vert ^{2}}\equiv F(\mathbf{\kappa })  \notag \\
=\sum\nolimits_{i,j=1}^{2}F_{ij}\kappa _{i}\kappa _{j}\ \ \left(
F_{ij}=F_{ji}\right)  \label{12}
\end{gather}%
\noindent\ It follows from (\ref{8}), (\ref{9}) and (\ref{11}), (\ref{12})
that
\begin{equation}
\mathbf{B}=\mu _{0}\mathbf{D}\left( \mathbf{I+C}\right) \mathbf{D},\mathbf{\
\ }M(\mathbf{\kappa })=\mu _{0}^{-1}\left( \left( \mathbf{I}+\mathbf{C}%
\right) ^{-1}\mathbf{f,f}\right) ,  \label{13}
\end{equation}%
where $\mathbf{I}$ is an infinite identity matrix. Note that $\mathbf{I+C}$
is positive and that it satisfies the identities%
\begin{multline}
\left( \mathbf{I}+\mathbf{C}\right) ^{-1}
=\sum\nolimits_{n=0}^{m}\left( - \mathbf{C}\right) ^{n}+\left(
-\mathbf{C}\right) ^{m+1}\left( \mathbf{I}+ \mathbf{C}\right) ^{-1},
\\
M(\mathbf{\kappa }) =\mu _{0}^{-1}\sum\nolimits_{n=0}^{m}\left(
\left( -
\mathbf{C}\right) ^{n}\mathbf{f,f}\right)  \\
+\mu _{0}^{-1}\left( \left( -\mathbf{C}\right) ^{m+1}\left(
\mathbf{I}+ \mathbf{C}\right) ^{-1}\mathbf{f,f}\right).\label{14}
\end{multline}
Taking (\ref{14}$_{2}$) with $m=0$ yields%
\begin{equation}
M(\mathbf{\kappa })=\mu _{0}^{-1}F(\mathbf{\kappa })-\mu _{0}^{-1}\left(
\mathbf{C}\left( \mathbf{I}+\mathbf{C}\right) ^{-1}\mathbf{f,f}\right) .
\label{15}
\end{equation}%
Consider (\ref{15}) for two different choices of $\mu _{0}>0$. If $\mu
_{0}=\max \mu (\mathbf{x})\equiv \mu _{\max }$ then $\mu _{\Delta }\left(
\mathbf{x}\right) =\mu (\mathbf{x})-\mu _{0}$ is negative, hence so is $%
\mathbf{C}$ and therefore the second term on the r.h.s. of (\ref{15}) is
positive. If $\mu _{0}=\min \mu \left( \mathbf{x}\right) \equiv \mu _{\min }$
then the above signs are inverted. Thus $\mu _{\max }^{-1}F(\mathbf{\kappa }%
)\leq M(\mathbf{\kappa })\leq \mu _{\min }^{-1}F(\mathbf{\kappa })$.
Combining this with (\ref{6}$_{2}$) gives the bounds%
\begin{equation}
\left\langle \mu \right\rangle -\frac{F(\mathbf{\kappa })}{\mu _{\min }}\leq
\mu _{\mathrm{eff}}(\mathbf{\kappa })\leq \left\langle \mu \right\rangle -%
\frac{F(\mathbf{\kappa })}{\mu _{\max }}\ \mathrm{for\ any}\ \mu (\mathbf{x}%
).  \label{16}
\end{equation}%
The lower bound is not very interesting since it may become negative if $\mu
_{\min }$ is small. The upper bound reinforces the inequality (\ref{10}) as
\begin{equation}
c^{2}(\mathbf{\kappa })\leq \frac{1}{\left\langle \rho \right\rangle }\left(
\left\langle \mu \right\rangle -\frac{F(\mathbf{\kappa })}{\mu _{\max }}%
\right) .  \label{17}
\end{equation}

It is natural to inquire as to what choice of $\mu _{0}$ provides the best
estimate of $\mu _{\mathrm{eff}}(\mathbf{\kappa })$ within the bounds (\ref%
{16}). To answer this question, let us formally consider Eqs. (\ref{14})
truncated as follows:%
\begin{gather}
\left( \mathbf{I}+\mathbf{C}\right) ^{-1}\approx
\sum\nolimits_{n=0}^{m}\left( -\mathbf{C}\right) ^{n},  \notag \\
M(\mathbf{\kappa })\approx \mu _{0}^{-1}\sum\nolimits_{n=0}^{m}\left( \left(
-\mathbf{C}\right) ^{n}\mathbf{f,f}\right) .  \label{17.1}
\end{gather}
The sufficient condition for convergence of both series as $m\rightarrow
\infty $ is $\left\Vert \mathbf{C}\right\Vert <1,$ where $\left\Vert \cdot
\right\Vert $ is an operator norm. Hence we need to take $\mu _{0}$ which
minimizes $\left\Vert \mathbf{C}(\mu _{0})\right\Vert $. Note from (\ref{12}%
) that $\mathbf{C}$ is close to the operator of multiplication by $\mu
_{\Delta }(\mathbf{x})/\mu _{0},$ so $\left\Vert \mathbf{C}\right\Vert $ may
be gauged by the value $\max_{\mathbf{x}}\left\vert \mu _{\Delta }\left(
\mathbf{x}\right) /\mu _{0}\right\vert $. Its minimum over all choices of $%
\mu _{0}$ is reached when $\mu _{0}=\frac{1}{2}\left( \mu _{\max }+\mu
_{\min }\right) $. Thus a simple estimate, given by a single first term $M(%
\mathbf{\kappa })\approx F(\mathbf{\kappa })/\mu _{0}$ of (\ref{17.1}$_{2}$%
), can be taken as%
\begin{gather}
c^{2}(\mathbf{\kappa })=\frac{\mu _{\mathrm{eff}}\left( \mathbf{\kappa }%
\right) }{\left\langle \rho \right\rangle }\approx \frac{1}{\left\langle
\rho \right\rangle }\left( \left\langle \mu \right\rangle -\frac{F(\mathbf{%
\kappa })}{\mu _{0}}\right)  \notag \\
\mathrm{with}\ \mu _{0}=\frac{\mu _{\max }+\mu _{\min }}{2}\equiv \overline{%
\mu }.  \label{18}
\end{gather}%
Note that the obtained estimation is a general result in the sense of having
the same form for an arbitrary periodic dependence $\mu \left( \mathbf{x}%
\right) ,$ but it certainly provides a different accuracy for different
types of $\mu (\mathbf{x})$. For instance, consider two extreme examples: a
stiff composite with small admixture of a highly contrasting soft ingredient
and the inverse case where these two components form a soft material with a
stiff reinforcement. The common ratio of geometrical progression (\ref{17.1}$%
_{2}$) with $\mu _{0}=\overline{\mu }$ has a similar absolute value (gauged
by $\max_{\mathbf{x}}\left\vert \mu _{\Delta }\left( \mathbf{x}\right) /%
\overline{\mu }\right\vert $) for both cases but is likely to differ in
sign, since $\mathbf{C}$ is close to multiplying by $\mu _{\Delta }\left(
\mathbf{x}\right) =\mu \left( \mathbf{x}\right) -\overline{\mu }$ and hence
should be positive (negative) definite when the stiff (respectively, soft)
component is volume dominant. Obviously a sign-alternating progression
converges faster. Thus the PWE estimate, which is the leading-order term of (%
\ref{17.1}$_{2}$), is expected to be more accurate in the former
case of a predominantly stiff composite with a small volume fraction
of a soft material and less accurate in the latter, inverse, case.
This observation is illuminated by the examples in \S
\ref{sec5.1.2}.

It remains to supply the closed-form relations for $F\left( \mathbf{\kappa }%
\right) $. From its definition in (\ref{12})$,$%
\begin{multline}
\mathrm{trace}\left( F_{ij}\right) =\sum\nolimits_{\mathbf{g}\in
\Gamma \backslash \left\{ \mathbf{0}\right\} }\left\vert
\widehat{\mu }\left(
\mathbf{g}\right) \right\vert ^{2}   \\
=\left\langle \left( \mu -\left\langle \mu \right\rangle \right)
^{2}\right\rangle =\left\langle \mu ^{2}\right\rangle -\left\langle
\mu \right\rangle ^{2}.  \label{19}
\end{multline}%
Hence by (\ref{17}) and (\ref{18}) the sum of squared effective speeds along
any pair of unit orthogonal vectors $\mathbf{\kappa }_{l}$ in $%
\mathbb{R}
^{2}$ satisfies%
\begin{eqnarray}
\sum\limits_{l=1}^{2}c^{2}\left( \mathbf{\kappa }_{l}\right) &\leq &\frac{1}{%
\left\langle \rho \right\rangle }\left( 2\left\langle \mu \right\rangle -%
\frac{\left\langle \mu ^{2}\right\rangle -\left\langle \mu \right\rangle ^{2}%
}{\mu _{\max }}\right) ,  \notag \\
\sum\limits_{l=1}^{2}c^{2}\left( \mathbf{\kappa }_{l}\right) &\approx &\frac{%
1}{\left\langle \rho \right\rangle }\left( 2\left\langle \mu \right\rangle -2%
\frac{\left\langle \mu ^{2}\right\rangle -\left\langle \mu \right\rangle ^{2}%
}{\mu _{\max }+\mu _{\min }}\right) .  \label{20}
\end{eqnarray}
The quadratic form $c^{2}(\mathbf{\kappa })$ is known to be independent of
the orientation of $\mathbf{\kappa }$ in $%
\mathbb{R}
^{2}$ if $\mu (\mathbf{x})$ (thus also $\widehat{\mu }\left( \mathbf{g}%
\right) $ and $c^{2}(\mathbf{\kappa })$) is invariant under three- or
fourfold rotations about the axis normal to the $\mathbf{x}$-plane.
In this case, $c^{2}(\mathbf{\kappa })=\frac{1}{2}\sum%
\nolimits_{l=1}^{2}c^{2}\left( \mathbf{\kappa }_{l}\right) $ for any $%
\mathbf{\kappa }$ and thus (\ref{20}) gives%
\begin{eqnarray}
c^{2} &\leq &\frac{1}{\left\langle \rho \right\rangle }\left( \left\langle
\mu \right\rangle -\frac{\left\langle \mu ^{2}\right\rangle -\left\langle
\mu \right\rangle ^{2}}{2\mu _{\max }}\right) ,  \notag \\
\ c^{2} &\approx &\frac{1}{\left\langle \rho \right\rangle }\left(
\left\langle \mu \right\rangle -\frac{\left\langle \mu ^{2}\right\rangle
-\left\langle \mu \right\rangle ^{2}}{\mu _{\max }+\mu _{\min }}\right)
\equiv c_{\mathrm{PWE}}^{2},  \label{21}
\end{eqnarray}
where the notation $c_{\mathrm{PWE}}^{2}$ is introduced for future use to
distinguish this estimate from those obtained by other methods. For a
piecewise homogeneous periodic material consisting of $J=1,2,..$. components
with constant $\mu _{J}$, $\rho _{J}$ and with filling fractions $f_{J}$ ($%
\sum f_{J}=1$), Eq. (\ref{21}) obviously specializes by setting $%
\left\langle \cdot \right\rangle =\sum_{J}\left( \cdot \right) _{J}f_{J}$
and $\mu _{\max /\min }=\left( \max /\min \right) _{J}\mu _{J}$.

The above results are formulated for the 2D periodic media; however, they
can be readily adapted for equations similar to (\ref{3}) with $\mathbf{x}%
\in
\mathbb{R}
^{d}$ of any dimension $d>2$, e.g., for 3D equations of heat conduction or
fluid acoustics. Indeed, replacing $\sum\nolimits_{l=1}^{2}$ by $%
\sum\nolimits_{l=1}^{d}$ keeps (\ref{19}) intact and replaces the factor 2
by $d$ before $\left\langle \mu \right\rangle $ in (\ref{20}), which leads
to $c^{2}(\mathbf{\kappa })=\frac{1}{d}\sum\nolimits_{l=1}^{d}c^{2}\left(
\mathbf{\kappa }_{l}\right) $ if $c^{2}\left( \mathbf{\kappa }\right) $ is
independent of $\mathbf{\kappa }\in
\mathbb{R}
^{d}$. This is the case for $d=3$ under cubic symmetry.
\footnote{Generally the condition for such isotropic behavior at any
$d>2$ may be stated as invariance of the coefficient $\mu
(\mathbf{x})$ in (\ref{3}) to the shift $x_{1}\rightarrow
x_{2},..,x_{d}\rightarrow x_{1}$ and, separately, to the change of
sign $x_{1}\rightarrow -x_{1}$ of the Cartesian coordinates $x_{i}$
of $\mathbf{x}\in \mathbb{R} ^{d}$ .}
For example, consider a
3D-periodic fluid-like cubic structure with bulk modulus
$K(\mathbf{x})$ and density $\rho (\mathbf{x})$. Based on the
standard equivalence between SH $\rightarrow $ acoustics under the
interchange $\rho \rightarrow K^{-1},$ $\mu \rightarrow \rho ^{-1},$
the PWE
bound and estimate of the effective acoustic speed follow in the form%
\begin{eqnarray}
c^{2} &\leq &\frac{1}{\left\langle K^{-1}\right\rangle }\left( \left\langle
\rho ^{-1}\right\rangle -\frac{\left\langle \rho ^{-2}\right\rangle
-\left\langle \rho ^{-1}\right\rangle ^{2}}{3\left( \rho ^{-1}\right) _{\max
}}\right) , \notag  \\
c^{2} &\approx &\frac{1}{\left\langle K^{-1}\right\rangle }\left(
\left\langle \rho ^{-1}\right\rangle -\frac{2}{3}\frac{\left\langle
\rho ^{-2}\right\rangle -\left\langle \rho ^{-1}\right\rangle
^{2}}{\rho _{\max }^{-1}+\rho _{\min }^{-1}}\right) . \label{22}
\end{eqnarray}

\subsection{MM approach and the estimate}

\label{sec3.2}

In this subsection we develop the $\mathbf{x}$-space approach basing on the
monodromy matrix (MM). The idea implies casting the wave equation in matrix
form containing an ordinary differential operator with quasi--periodic
boundary condition in one coordinate, integrating this system using the
multiplicative integral in the other coordinate, and applying perturbation
theory to express the result via the scalar product in $L^{2}\left( \mathbf{T%
}\right) $ that enable eliminating the operators and yields the closed-form
approximate solution in the form of double integrals of $\mu (\mathbf{x})$
and $\rho (\mathbf{x})$. Thus the MM approach is performed in $\mathbf{x}$%
-space.

It is convenient to assume for the moment that the functions $\mu (\mathbf{x}%
)$ and $\rho (\mathbf{x})$ in the wave equation (\ref{3}) are smooth
functions, which are periodic on the 2D rectangular lattice with the
unit cell $\mathbf{T\ni x}=\left( x_{1},x_{2}\right) $ formed by the
translations $\mathbf{a}_{1,2}\parallel \mathbf{e}_{1,2}$ (see \S
\ref{sec2}). Alongside the notation $\left\langle \cdot
\right\rangle \equiv \frac{1}{\left\vert \mathbf{T}\right\vert
}\int_{\mathbf{T}}\cdot ~\mathrm{d}\mathbf{x}$
introduced in (\ref{4}), denote%
\begin{equation}
\left\langle \cdot \right\rangle _{x_{i}}\equiv \frac{1}{\left\vert \mathbf{a%
}_{i}\right\vert }\int_{0}^{\left\vert \mathbf{a}_{i}\right\vert }\cdot ~%
\mathrm{d}x_{i}\ \ \ \left( \Rightarrow \left\langle \left\langle \cdot
\right\rangle _{x_{1}}\right\rangle _{x_{2}}=\left\langle \cdot
\right\rangle \right)   \label{M1}
\end{equation}%
and let, for brevity, $\mathbf{a}_{1,2}$ be of unit length so that $\mathbf{T%
}=\left[ 0,1\right] ^{2}$. Imposing the Floquet quasi-periodic condition
along one of the coordinates, say $x_{1}$, leads to $v\left( \mathbf{x}%
\right) =w\left( \cdot ,x_{2}\right) e^{ik_{1}x_{1}}$ where $w\left( \cdot
,x_{2}\right) \equiv w\left( x_{1}\right) $ for any fixed $x_{2}$ and $%
w\left( x_{1}\right) $ is an absolutely continuous periodic function:%
\begin{eqnarray}
w\left( x_{1}\right)  &\in &W\equiv \left\{ w\left( x_{1}\right) \in AC\left[
0,1\right] :\right. \   \notag \\
&&\ \left. w\left( 0\right) =w\left( 1\right) ,\ w^{\prime }\left( 0\right)
=w^{\prime }\left( 1\right) \right\}   \label{M2}
\end{eqnarray}%
with $^{\prime }$ meaning $\mathrm{d}/\mathrm{d}x_{1}$. On these grounds,
Eq. (\ref{3}) can be rewritten in the form%
\begin{gather}
\mathcal{Q}\mathbf{\eta }=\frac{\partial }{\partial x_{2}}\mathbf{\eta },\
\mathcal{Q}=\left(
\begin{array}{cc}
0 & \mu ^{-1}(\mathbf{x}) \\
\mathcal{A}-\omega ^{2}\rho (\mathbf{x}) & 0%
\end{array}%
\right) ,  \notag \\
\mathbf{\eta }(\mathbf{x})=\left(
\begin{array}{c}
w\left( \cdot ,x_{2}\right)  \\
\mu (\mathbf{x})\partial w\left( \cdot ,x_{2}\right) /\partial x_{2}%
\end{array}%
\right) ,  \label{M3}
\end{gather}%
where the operator $\mathcal{A=A}\left( k_{1},x_{2}\right) $ acting on the
components of $\mathbf{\eta }$ as on functions of $x_{1}$ is defined in the
space $W$ by the definition%
\begin{multline}
\mathcal{A}\left( k_{1},x_{2}\right) w\left( x_{1}\right)\\
=-e^{-ik_{1}x_{1}}\left( \mu \left( x_{1},\cdot \right) \left(
e^{ik_{1}x_{1}}w\left( x_{1}\right) \right) ^{\prime }\right)
^{\prime }
\\
=-\left( \mu w^{\prime }\right) ^{\prime }-ik_{1}\left( \mu
w^{\prime }+\left( \mu w\right) ^{\prime }\right) +k_{1}^{2}\mu
w.\label{M4}
\end{multline}%
The solution $\mathbf{\eta }\left( \mathbf{x}\right) $ of Eq. (\ref{M3})
with the initial condition $\mathbf{\eta }\left( x_{1},0\right) =\mathbf{%
\eta }_{0}\left( x_{1}\right) $ at $x_{2}=0$ can be represented in the form%
\begin{gather}
\mathbf{\eta }\left( \mathbf{x}\right) =\mathcal{M}\left[ x_{2},0\right]
\mathbf{\eta }_{0}\left( x_{1}\right) \ \mathrm{with}  \notag \\
\mathcal{M}\left[ x_{2},0\right] =\widehat{\int }_{0}^{x_{2}}\left( \mathcal{%
I}+\mathcal{Q}\mathrm{d}x_{2}\right) =\mathcal{I}+\int_{0}^{x_{2}}\mathcal{Q}%
\left( k_{1},\varsigma \right) \mathrm{d}\varsigma   \notag \\
+\int_{0}^{x_{2}}\mathcal{Q}\left( k_{1},\varsigma \right) \mathrm{d}%
\varsigma \int_{0}^{\varsigma }\mathcal{Q}\left( k_{1},\varsigma
_{1}\right) \mathrm{d}\varsigma _{1}+...,\label{M5}
\end{gather}%
where $\mathcal{I}$ is the identity operator, and the operator
$\mathcal{M}\left[ x_{2},0\right] $ is formally a matricant of
(\ref{M3}) defined in a standard fashion through a
multiplicative integral $\widehat{\int }$ expanding in the Peano series \cite%
{P}. In the same spirit, the operator $\mathcal{M}\left[ 1,0\right] $ given
by (\ref{M5}) with $x_{2}=1,$ i.e. taken over a period 1 in $x_{2}$, may be
called a monodromy matrix. It has the important property that if $%
e^{ik_{2}\left( \omega ,k_{1}\right) }$ with $k_{2}\in
\mathbb{R}
$ is an eigenvalue of $\mathcal{M}\left[ 1,0\right] $, then $\omega $ and $%
\mathbf{k}=\left( k_{1},k_{2}\right) $ satisfy Eq. (\ref{3}), i.e. $\omega
^{2}$ is an eigenvalue of (\ref{3}) with the Floquet quasi-periodic
conditions along both coordinates $x_{1}$ and $x_{2}$. This is similar to
the case of scalar waves in 2D media with 1D periodicity (see \cite{KSNP});
however, the presence of terms of the order $O\left( k_{1}^{0}\right) ,$ $%
O\left( k_{1}\right) \ni \mathcal{A}$ in (\ref{M4}) underlies an essential
difference in the 2D periodicity case. Note that $\mathcal{M}\left[ a,b%
\right] $ at $\omega =0,~k_{1}=0$ has the eigenvalue $e^{ik_{2}\left(
0,0\right) }=1$ corresponding to the eigenvector $\left( 1,0\right) ^{%
\mathrm{T}},$ i.e., to $w\left( x_{1}\right) =const$.

The MM approach enables deriving a new form of the exact solution for the
effective speed. Referring for brevity to the isotropic case, it is as
follows:
\begin{equation}
c^{2}=\frac{1}{\langle \rho \rangle }\left\langle \left( 0,1\right)
\left( \mathcal{M}_{1}\left[ 1,0\right] -\mathcal{I}\right)
^{-1}\left( 1,0\right) ^{\mathrm{T}}\right\rangle _{x_{1}},
\label{503}
\end{equation}%
where $\mathcal{M}_{1}\left[ 1,0\right] $ is $\mathcal{M}\left[ 1,0\right] $
with $\omega ,~k_{1}=0$ and $\left( \mathcal{M}_{1}\left[ 1,0\right] -%
\mathcal{I}\right) ^{-1}\left( 1,0\right) ^{\mathrm{T}}$ is any vector from
the preimage of the vector $\left( 1,0\right) ^{\mathrm{T}}$ with respect to
$\mathcal{M}_{1}\left[ 1,0\right] -\mathcal{I}$. We will not, however,
discuss Eq. (\ref{503}) in detail because, as any exact solution for $c$, it
defies a closed form and hence exceeds the scope of the present study.

Seeking specifically a closed-form estimate of $c$ necessitates some
additional simplifications. On this ground, let us further consider the
matrix operator $\mathcal{M}_{0}$ which consists of the first two terms of
the Peano series of $\mathcal{M}\left[ 1,0\right] $ (see (\ref{M5}) with $%
x_{2}=1$):%
\begin{equation}
\mathcal{M}\left[ 1,0\right] =\mathcal{M}_{0}+\ldots \quad \mathrm{with}\
\mathcal{M}_{0}\left( \omega ,k_{1}\right) =\mathcal{I}+\left\langle
\mathcal{Q}\right\rangle _{x_{2}}.  \label{M7}
\end{equation}%
Denote by $e^{i\widetilde{k}_{2}}$ and $\mathbf{e}$ the eigenvalue and
eigenvector of $\mathcal{M}_{0}$ which at $\omega =0,~k_{1}=0$ coincide with
those of $\mathcal{M}\left[ 1,0\right] ,$ so that%
\begin{gather}
\mathcal{M}_{0}\mathbf{e}\left( \omega ,k_{1};x_{1}\right) =e^{i\widetilde{k}%
_{2}\left( \omega ,k_{1}\right) }\mathbf{e}\left( \omega ,k_{1};x_{1}\right)
,  \notag \\
\mathrm{where}\ \widetilde{k}_{2}\left( 0,0\right) =0,\ \mathbf{e}\left(
0,0;x_{1}\right) =\left( 1,0\right) ^{\mathrm{T}}.  \label{M8}
\end{gather}%
The motivation for introducing $\mathcal{M}_{0}$ is that $\widetilde{k}%
_{2}\left( \omega ,k_{1}\right) $ has an exact closed-form asymptotic form
that can be used for constructing an estimate of $c$. It is emphasized that
the difference between $\mathcal{M}\left[ 1,0\right] $ and $\mathcal{M}_{0}$%
, which is given by the members of the Peano series (\ref{M5}) of the order $%
n>2$, contains the terms of the same order $O\left( k_{1}^{0}\right) ,$ $%
O\left( k_{1}\right) $ ($\ni \mathcal{A}$) and $O\left( \omega ^{2}\right) $
as in $\mathcal{M}_{0}$ but with numerical factors decreasing somewhat like $%
1/n!$. For the latter reason, the asymptotics of $\widetilde{k}_{2}\left(
\omega ,k_{1}\right) $ and $k_{2}\left( \omega ,k_{1}\right) $ should be
close.

To obtain the asymptotics of $\widetilde{k}_{2}\left( \omega ,k_{1}\right) $
in small $\omega ,$ $k_{1},$ it is convenient to pass from the matrix form
of (\ref{M5}) to the scalar equation as follows:%
\begin{gather}
\left\langle \mathcal{Q}\right\rangle _{x_{2}}\mathbf{e}=\lambda \mathbf{e}%
\mathrm{\ }\left( \lambda \equiv e^{i\widetilde{k}_{2}}-1\right) \Rightarrow
\notag \\
\left\langle \mu ^{-1}\right\rangle _{x_{2}}\left( \left\langle \mathcal{A}%
\right\rangle _{x_{2}}-\omega ^{2}\left\langle \rho \right\rangle
_{x_{2}}\right) e_{1}=\lambda ^{2}e_{1},  \label{M9}
\end{gather}%
where $\lambda =0$ and $e_{1}\left( x_{1}\right) =1$ at $\omega =0,~k_{1}=0$
by (\ref{M8}). Denote $k_{1}=\alpha \varepsilon ~$and $\omega =\beta
\varepsilon $ where $\varepsilon $ is a small perturbation parameter.
Inserting in (\ref{M9})$_{3}$ and invoking (\ref{M4}) yields%
\begin{gather}
\left( \mathcal{R}_{0}+\varepsilon \mathcal{R}_{1}+\varepsilon ^{2}\mathcal{R%
}_{2}\right) e_{1}\left( \varepsilon ,x_{1}\right) =\lambda ^{2}\left(
\varepsilon \right) \mathcal{D}e_{1}\left( \varepsilon ,x_{1}\right) \ \
\notag \\
\mathrm{with}\ \ \mathcal{D}w=\left\langle \mu ^{-1}\right\rangle
_{x_{2}}^{-1}w,\ \mathcal{R}_{0}w\equiv -\left( \left\langle \mu
\right\rangle _{x_{2}}w^{\prime }\right) ^{\prime },  \notag \\
\ \mathcal{R}_{1}w\equiv -i\alpha \left( \left\langle \mu \right\rangle
_{x_{2}}w^{\prime }+\left( \left\langle \mu \right\rangle _{x_{2}}w\right)
^{\prime }\right) ,  \notag \\
\mathcal{R}_{2}w\equiv \left( \alpha ^{2}\left\langle \mu \right\rangle
_{x_{2}}-\beta ^{2}\left\langle \rho \right\rangle _{x_{2}}\right) w.
\label{M10}
\end{gather}%
Note that the operators $\mathcal{R}_{i}$ and $\mathcal{D}$ acting on $%
w\left( x_{1}\right) \in W$ are self-adjoint with respect to the inner
product $\left( f,h\right) \equiv \int_{0}^{1}fh^{\ast }\mathrm{d}%
x_{1}\equiv \left\langle fh^{\ast }\right\rangle _{x_{1}}$ where $^{\ast }$
means complex conjugate. Applying the standard technique of perturbation
theory then leads to 
\begin{eqnarray}
\lambda ^{2}(\varepsilon ) &=&(\lambda ^{2})_{1}\varepsilon +(\lambda
^{2})_{2}\varepsilon ^{2}+O(\varepsilon ^{3})\text{ }\mathrm{with}  \notag \\
\left( \lambda ^{2}\right) _{1} &=&\frac{\left( \mathcal{R}%
_{1}e_{01},e_{01}\right) }{\left( \mathcal{D}e_{01},e_{01}\right) },  \notag
\\
\left( \lambda ^{2}\right) _{2} &=&\frac{\left( \mathcal{R}%
_{2}e_{01},e_{01}\right) -\left( \mathcal{R}_{0}^{-1}\mathcal{R}_{1}e_{01},%
\mathcal{R}_{1}e_{01}\right) }{\left( \mathcal{D}e_{01},e_{01}\right) },
\label{M11}
\end{eqnarray}%
where $e_{01}\equiv e_{1}(0,x_{1})=1$. First note that $\left( \mathcal{R}%
_{1}e_{01},e_{01}\right) =-i\alpha \int\nolimits_{0}^{1}\left( \left\langle
\mu \right\rangle _{x_{2}}\right) ^{\prime }\mathrm{d}x_{1}=0$ since $%
\left\langle \mu \right\rangle _{x_{2}}$ is a periodic function of $x_{1}$,
and hence $\left( \lambda ^{2}\right) _{1}=0$. To find $\mathcal{R}_{0}^{-1}%
\mathcal{R}_{1}e_{01}\equiv \phi \left( x_{1}\right) $, we need to solve the
equation $\mathcal{R}_{0}\phi \left( x_{1}\right) =\mathcal{R}_{1}e_{01},$
that is,%
\begin{gather}
-\left( \left\langle \mu \right\rangle _{x_{2}}\phi ^{\prime }\right)
^{\prime }=-i\alpha \left( \left\langle \mu \right\rangle _{x_{2}}\right)
^{\prime }\ \ \ \Rightarrow   \notag \\
\ \phi \left( x_{1}\right) =K+i\alpha
x_{1}+K_{1}\int_{0}^{x_{1}}\left\langle \mu \right\rangle _{x_{2}}^{-1}%
\mathrm{d}x_{1},  \label{M12}
\end{gather}%
where $K$ and $K_{1}$ are constants. Using the boundary condition $\phi
(0)=\phi (1)$ for $\phi (x_{1})\in W$ (see (\ref{M2})) determines $K_{1},$
whence%
\begin{eqnarray}
\mathcal{R}_{0}^{-1}\mathcal{R}_{1}e_{01} &\equiv &\phi \left( x_{1}\right)
=K+i\alpha x_{1}  \notag \\
&&-i\alpha \left\langle \left\langle \mu \right\rangle
_{x_{2}}^{-1}\right\rangle _{x_{1}}^{-1}\int_{0}^{x_{1}}\left\langle \mu
\right\rangle _{x_{2}}^{-1}\mathrm{d}x_{1}.  \label{M13}
\end{eqnarray}%
Thus, calculating%
\begin{eqnarray}
\left( \mathcal{R}_{0}^{-1}\mathcal{R}_{1}e_{01},\mathcal{R}%
_{1}e_{01}\right)  &=&\alpha ^{2}\left[ \left\langle \left\langle \mu
\right\rangle _{x_{2}}\right\rangle _{x_{1}}-\left\langle \left\langle \mu
\right\rangle _{x_{2}}^{-1}\right\rangle _{x_{1}}^{-1}\right] ,  \notag \\
\ \left( \mathcal{R}_{2}e_{01},e_{01}\right)  &=&\alpha ^{2}\left\langle
\left\langle \mu \right\rangle _{x_{2}}\right\rangle _{x_{1}}-\beta
^{2}\left\langle \rho \right\rangle ,  \notag \\
\ \left( \mathcal{D}e_{01},e_{01}\right)  &=&\left\langle \left\langle \mu
^{-1}\right\rangle _{x_{2}}^{-1}\right\rangle _{x_{1}},  \label{M14}
\end{eqnarray}%
and inserting in (\ref{M11}) yields the explicit form of $\lambda
^{2}(\varepsilon )\equiv \big(e^{i\widetilde{k}_{2}}-1\big)^{2}\approx
\left( \lambda ^{2}\right) _{2}\varepsilon ^{2}$ which, on reverting to the
original parameters $k_{1}=\alpha \varepsilon $ and $\omega =\beta
\varepsilon \mathbf{,}$ yields%
\begin{equation}
-\widetilde{k}_{2}^{2}=\frac{k_{1}^{2}\left\langle \left\langle \mu
\right\rangle _{x_{2}}^{-1}\right\rangle _{x_{1}}^{-1}-\omega
^{2}\left\langle \rho \right\rangle }{\left\langle \left\langle \mu
^{-1}\right\rangle _{x_{2}}^{-1}\right\rangle _{x_{1}}}+O\left(
k_{1}^{3},\omega ^{3}\right) .  \label{M15}
\end{equation}%
As argued above, $\widetilde{k}_{2}\left( \omega ,k_{1}\right) $ at small $%
\omega $ and $k_{1}$ is supposed to be close to $k_{2}\left( \omega
,k_{1}\right) $; therefore replacing $\widetilde{k}_{2}$ in (\ref{M15}) by $%
k_{2}$ leads to the approximation for the effective speed $c\left( \mathbf{%
\kappa }\right) =\lim_{\omega ,k\rightarrow 0}\omega \left( \mathbf{k}%
\right) /k$ ($\mathbf{k}=k\mathbf{\kappa }$) as%
\begin{equation}
c^{2}(\mathbf{\kappa })\approx \frac{1}{\left\langle \rho \right\rangle }%
\left( \kappa _{1}^{2}\left\langle \left\langle \mu \right\rangle
_{x_{2}}^{-1}\right\rangle _{x_{1}}^{-1}+\kappa _{2}^{2}\left\langle
\left\langle \mu ^{-1}\right\rangle _{x_{2}}^{-1}\right\rangle
_{x_{1}}\right) .  \label{M16}
\end{equation}

Note that applying the same scheme with respect to the reverse order of
coordinates, i.e. imposing the Floquet condition along $x_{2}$ and using the
monodromy matrix along $x_{1}$, yields the formula which follows from (\ref%
{M16}) by interchanging $x_{1}\rightleftarrows x_{2}\ $and $\kappa
_{1}\rightleftarrows \kappa _{2}$. Neither of the two approximations is
generally preferable, so it is natural to use their average, say, the
half-sum $\frac{1}{2}\left[ (\ref{M16})+(\ref{M16})_{1\rightleftarrows 2}%
\right] $. Thereby we arrive at the estimate for the effective speed in the
following form:%
\begin{eqnarray}
c^{2}(\mathbf{\kappa }) &\approx &\frac{1}{2\left\langle \rho \right\rangle }%
\left[ \left( \left\langle \left\langle \mu ^{-1}\right\rangle
_{x_{1}}^{-1}\right\rangle _{x_{2}}+\left\langle \left\langle \mu
\right\rangle _{x_{2}}^{-1}\right\rangle _{x_{1}}^{-1}\right) \kappa
_{1}^{2}\right.   \notag \\
&&\left. +\left( \left\langle \left\langle \mu \right\rangle
_{x_{1}}^{-1}\right\rangle _{x_{2}}^{-1}+\left\langle \left\langle \mu
^{-1}\right\rangle _{x_{2}}^{-1}\right\rangle _{x_{1}}\right) \kappa _{2}^{2}
\right] ,  \label{M17}
\end{eqnarray}
where $\left\langle \cdot \right\rangle _{x_{i}}$ is defined by (\ref{M1})
(obviously the assumption of unit and equal periods $\left\vert \mathbf{a}%
_{i}\right\vert $ is no longer needed).

The wave speed estimate (\ref{M17}) describes an ellipse of effective
slowness $\mathbf{s}(\mathbf{\kappa })=c^{-1}\left( \mathbf{\kappa }\right)
\mathbf{\kappa }$ with the principal axes along the translations $\mathbf{a}%
_{1}\perp \mathbf{a}_{2}$ of an orthotropic lattice. For an isotropic
lattice, where each of $\left\langle \left\langle \mu \right\rangle
_{x_{1}}^{-1}\right\rangle _{x_{2}}^{-1}\neq \left\langle \left\langle \mu
^{-1}\right\rangle _{x_{1}}^{-1}\right\rangle _{x_{2}}$ is invariant to $%
x_{1}\rightleftarrows x_{2},$ Eq. (\ref{M17}) (in contrast to (\ref{M16}))
becomes isotropic, i.e., yields the same value%
\begin{equation}
c_{\mathrm{MM}}^{2}=\frac{1}{2\left\langle \rho \right\rangle }\left(
\left\langle \left\langle \mu ^{-1}\right\rangle _{x_{1}}^{-1}\right\rangle
_{x_{2}}+\left\langle \left\langle \mu \right\rangle
_{x_{2}}^{-1}\right\rangle _{x_{1}}^{-1}\right)   \label{M18}
\end{equation}%
for any $\mathbf{\kappa .}$ It is instructive to apply the explicit formula (%
\ref{M18}) to a square lattice composed of $J=1,2,..$. homogeneous
materials, which is the case exemplified in detail in \S \ref{sec5}.
Inserting $\mu \left( \mathbf{x}\right) =\sum_{J}\mu _{J}\chi
_{J}\left( \mathbf{x}\right) $
for $\mathbf{x\in T,}$ where $\chi _{J}\left( \mathbf{x}\right) $ ($%
\left\langle \chi _{J}\right\rangle =f_{J}$) is an indicator function equal
to 1 on the domain occupied by the $J^{\text{th}}$ material and to 0
elsewhere, reduces Eq. (\ref{M18}) to%
\begin{eqnarray}
c_{\mathrm{MM}}^{2} &=&\frac{1}{2\left\langle \rho \right\rangle }\left[
\int_{0}^{1}\frac{\mathrm{d}\varsigma _{2}}{\sum_{J}\mu _{J}^{-1}\chi
_{J}\left( \varsigma _{2}\right) }\right.   \notag \\
&&\left. +\left( \int_{0}^{1}\frac{\mathrm{d}\varsigma _{2}}{\sum_{J}\mu
_{J}\chi _{J}\left( \varsigma _{2}\right) }\right) ^{-1}\right]
\label{M18.1}
\end{eqnarray}
with $\chi _{J}\left( \varsigma _{2}\right) =\int_{0}^{1}\chi _{J}\left(
\varsigma _{1},\varsigma _{2}\right) \mathrm{d}\varsigma _{1}$ and $%
\varsigma _{i}=x_{i}/\left\vert \mathbf{a}_{i}\right\vert $. Now suppose
that one of the constituent materials has $\mu _{J}\rightarrow 0$ and it is
distributed with a small (but finite) concentration $f_{J}$ along the
unit-cell boundary. Then both integrals on the r.h.s. of (\ref{M18.1}) tend
to zero, and so $c_{\mathrm{MM}}^{2}\rightarrow 0.$ Thus the essential
attribute of the MM estimate (\ref{M18.1}) is that it is capable of
capturing the 'insulating' effect of even a small concentration of soft
material when this forms a 'network' breaking the connectivity of stiff
components in the lattice. One more revealing example is the limiting case
where $\mu (\mathbf{x})$ is constant along some fixed direction in $%
\mathbb{R}
^{2}$ (while $\rho (\mathbf{x})$ may remain 2D-periodic). Taking this
direction as the base vector $\mathbf{e}_{1}$ implies $\left\langle \mu
\right\rangle _{x_{1}}=\mu \left( x_{2}\right) $ and thus reduces (\ref{M17}%
) to the well-known exact formula
\begin{equation}
c^{2}(\mathbf{\kappa })=\left\langle \rho \right\rangle ^{-1}\left(
\left\langle \mu \right\rangle _{x_{2}}\kappa _{1}^{2}+\left\langle \mu
^{-1}\right\rangle _{x_{2}}^{-1}\kappa _{2}^{2}\right) .  \label{M19}
\end{equation}%
In fact, the original non-symmetric estimate (\ref{M16}) reduces to the
exact form (\ref{M19}) when $\mu (\mathbf{x})$ is constant along the
direction $\mathbf{e}_{i}$, $i=1$ or $2$.

In conclusion, a few remarks are in order concerning the approximate nature
of the MM-approach implementation and result. First, the assumption that $%
\mu (\mathbf{x})$ and $\rho (\mathbf{x})$ are smooth can actually be
relaxed to include piecewise continuous functions and hence to apply
the approximation (\ref{M17}) to composites with inclusions, see \S
\ref{sec5}. This is
similar to the effect of truncating PWE series of piecewise continuous $\mu (%
\mathbf{x})$ and $\rho \left( \mathbf{x}\right) $, which allows one to think
of them as smooth functions (\S \ref{sec4}). A second remark is that the MM estimate (%
\ref{M17}) is not restricted to the isotropic case like the PWE estimate (%
\ref{21}$_{2}$) is. On the other hand, due to the simplification adopted on
deriving Eq. (\ref{M17}), it does not contain a cross term proportional to $%
\kappa _{1}\kappa _{2}$ and hence is unable to pinpoint the effect of
asymmetric form and/or distribution of inclusions in a rectangular lattice
that could tilt the principal axes of the exact effective-speed curve away
from the translation vectors $\mathbf{a}_{1},~\mathbf{a}_{2}$. For the same
reason, Eq. (\ref{M17}) may not be invariant with respect to different
choices of a unit cell in a given lattice. Such deficiency could be
rectified by taking into account the terms of order $O\left( \omega
^{2}\right) $ from the next ($n>2$) Peano-series members, which are
discarded in $\mathcal{M}_{0}$ (cf. (\ref{M5}) and (\ref{M7})); however,
this is hardly an expedient course of action since adding even one more term
on top of $\mathcal{M}_{0}$ leads to quite a cumbersome expression for $c$.
Finally, we note that instead of taking the arithmetic mean leading to (\ref%
{M17}), one could have invoked another average, e.g., the geometric mean.
Its direct use as $\left[ (\ref{M16})\times (\ref{M16})_{1\rightleftarrows 2}%
\right] ^{1/2}$ is unreasonable since the resulting estimate of squared
speed $c^{2}(\mathbf{\kappa })$ would no longer be a quadratic form in $%
\mathbf{\kappa ;}$ however, the geometric mean could be applied separately
to the coefficients of $\kappa _{j}$ thus yielding%
\begin{gather}
c^{2}(\mathbf{\kappa })\approx \frac{1}{\left\langle \rho \right\rangle }%
\left[ \left( \left\langle \left\langle \mu ^{-1}\right\rangle
_{x_{1}}^{-1}\right\rangle _{x_{2}}\left\langle \left\langle \mu
\right\rangle _{x_{2}}^{-1}\right\rangle _{x_{1}}^{-1}\right) ^{1/2}\kappa
_{1}^{2}\right.   \notag \\
\ \ \ \ \ \ \ \left. +\left( \left\langle \left\langle \mu
^{-1}\right\rangle _{x_{2}}^{-1}\right\rangle _{x_{1}}\left\langle
\left\langle \mu \right\rangle _{x_{1}}^{-1}\right\rangle
_{x_{2}}^{-1}\right) ^{1/2}\kappa _{2}^{2}\right] \equiv c_{\widetilde{%
\mathrm{MM}}}^{2}.  \label{M20}
\end{gather}
Comparison of the two MM estimates $c_{\widetilde{\mathrm{MM}}}$\ and $c_{%
\mathrm{MM}}$\ is considered in \S \ref{sec5.1.1}.

\section{PWE numerical implementation}

\label{sec4}

PWE numerical implementation rests on calculation of the quantity $M(\mathbf{%
\kappa })=\left( \mathbf{B}^{-1}\mathbf{d,d}\right) ,$ Eq. (\ref{8}), which
involves the inverse of the formally infinite matrix $\mathbf{B}$ truncated
in the 2D calculations to a finite $N^{2}\times N^{2}$ size ($N$ is the
number of Fourier terms in one coordinate). Its inversion takes $O\left(
N^{8}\right) $ steps. Calculating $\mathbf{B}^{-1}\mathbf{d,}$ i.e. solving
a linear system $\mathbf{Bh=d}$ for unknown $\mathbf{h}$ by Gauss or similar
methods, takes $O\left( N^{6}\right) $ steps (and needs $O\left(
N^{4}\right) $ memory cells for storing intermediate results). This may also
be onerous for large enough $N$. Note also that the case of high-contrast
lattices with very soft or void components needs special care (see e.g. \cite%
{V-H}). The difficulty arises due to the fact that $\mathbf{B}$ is not
invertible if $\mu \left( \mathbf{\Omega }\right) =0$ for some domain $%
\mathbf{\Omega }$ of $\mathbf{x}$ within the unit cell $\mathbf{T}$. This
does not preclude numerical inversion of truncated $\mathbf{B}$ (since a
finite-size $\mathbf{B}$ can no longer possess eigenfunctions with a support
in $\mathbf{\Omega }$ $\subsetneqq $ $\mathbf{T}$); however, both inversion
of $\mathbf{B}$ and solving $\mathbf{Bh=d}$ with zero or small $\mu \left(
\mathbf{\Omega }\right) $ may become tricky because taking more elements of $%
\mathbf{B}$ implies a greater impact of its small eigenvalues and thus may
actually deteriorate numerical accuracy.

In this light, we advocate the method of direct computation of $M\left(
\mathbf{\kappa }\right) $ via the series expansion (\ref{17.1}$_{2}$) with $%
\mu _{0}=\frac{1}{2}\left( \mu _{\max }+\mu _{\min }\right) \equiv \overline{%
\mu }.$ On fixing the meaning of truncated quantities as defined on a $N^{2}$%
-dimension subspace $l_{N^{2}}^{2}\subset l^{2}\left( \Gamma \backslash
\left\{ \mathbf{0}\right\} \right) $ spanned by $N^{2}=\left( 2j+1\right)
^{2}$ vectors $\mathbf{e}_{\mathbf{g}}=\left( \delta _{\mathbf{gg}^{\prime
}}\right) _{\mathbf{g}^{\prime }\neq \mathbf{0}}$ with $0<\left\vert
g_{i}\right\vert \leq 2\pi j$ ($i=1,2$), the explicit expression for
computing $M\left( \mathbf{\kappa }\right) $ is
\begin{equation}
M(\mathbf{\kappa })\approx \overline{\mu }^{-1}\sum\nolimits_{n=0}^{m}\left(
\left( -\mathbf{C}_{N^{2}\times N^{2}}\right) ^{n}\mathbf{f}_{N^{2}}\mathbf{%
,f}_{N^{2}}\right) ,  \label{17.2}
\end{equation}%
where $\mathbf{C}_{N^{2}\times N^{2}}(\overline{\mu })\equiv \mathbf{\mathbf{%
\mathbf{C}}}$ and $\mathbf{f}_{N^{2}}\equiv \mathbf{f}$ have components $%
\left( \mathbf{Ce}_{\mathbf{g}},\mathbf{e}_{\mathbf{g}^{\prime }}\right) $
and $\left( \mathbf{f},\mathbf{e}_{\mathbf{g}}\right) $ in $l_{N^{2}}^{2}$.
'Termwise' (by way of storing $\mathbf{\mathbf{\mathbf{C}}}^{n}\mathbf{f}$
and calling on it for $\mathbf{C}^{n+1}\mathbf{f=\mathbf{C}}\left( \mathbf{%
\mathbf{\mathbf{C}}}^{n}\mathbf{f}\right) $) calculation of (\ref{17.2})
takes $O\left( mN^{4}\right) $ steps, which is notably less than $O\left(
N^{6}\right) $ when $N\gg m,1$. The validity of approximation (\ref{17.2})
can be justified on the basis of the sufficient condition $\left\Vert
\mathbf{C}\right\Vert <1$ for convergence of (\ref{17.1}$_{1}$) and on the
fact that $\mathbf{C}$ is close to the operator of multiplication by $(\mu (%
\mathbf{x})-\mu _{0})/\mu _{0}$ whence $\left\Vert \mathbf{C}(\overline{\mu }%
)\right\Vert \sim \left\vert \mu (\mathbf{x})/\overline{\mu }-1\right\vert $
(see the discussion of Eqs. (\ref{17.1}), (\ref{18}) in \S \ref{sec3}). Thus $%
\left\Vert \mathbf{C}(\overline{\mu })\right\Vert $ is expected to
be less than 1, being close to 1 in the special case where $\mu $ is
very small in some $\mathbf{\Omega }\in \mathbf{T.}$ In the former
case, fast convergence of (\ref{17.1}$_{2}$) is facilitated by the
diagonal predominant structure of $\mathbf{I+C}$ (see Appendix). In
the latter case (small $\mu \left( \mathbf{\Omega }\right) $), the
fact that $\left\vert \mathbf{f}\right\vert $ decreases as
$\mathbf{g}$ grows large may come into play. However, in contrast to
the MM approach (see \S \ref{sec3.2}), the PWE considerations seem
unable to explain the very different effect of this small $\mu $
when it occurs either strictly inside the unit cell (soft inclusion)
or along its
boundaries (soft matrix). Numerical examples provided in \S \ref{sec5} show that Eq. (%
\ref{17.2}) is not sensitive to $\mu $ of an inclusion tending to
zero, and hence it can be directly applied to computing the
effective shear speed in solid/air and solid/fluid composites (see
e.g. \cite{GV,S-D}), where the solid phase remains
connected\footnote{{\small Note that the effective density for shear
(SH) waves in 2D solid-fluid structures depends only on the solid
density since the vanishing shear force on the fluid/solid interface
means that the fluid does not participate in the SH motion.}}. The
alternative case, in which a very soft matrix material forms an
'insulating network', is known to be particularly subtle for any
PWE-based numerical scheme. No wonder that application of Eq.
(\ref{17.2}) to this case requires more numerical effort as detailed
in \S \ref{sec5}.

Note that taking (\ref{17.1}$_{2}$) with $\mu _{0}=\overline{\mu },$ which
leads to the same form (\ref{17.2}) for any $\mu (\mathbf{x})$, does not at
all guarantee the fastest convergence for all $\mu (\mathbf{x})$. This is
elucidated in Appendix which contains an example of strict and quantitative
convergence analysis of the series (\ref{17.1}$_{2}$) for a particular
family of periodic $\mu (\mathbf{x})$.

\section{Discussion and examples}

\label{sec5}

\subsection{Two-phase lattices}

\label{sec5.1}

\subsubsection{Estimates}

\label{sec5.1.1}

Consider a 2D square lattice which is isotropically composed of two
homogeneous materials $J=1,2$ with constant $\rho _{J},$ $\mu _{J}$
and with filling fractions $f_{J}$ ($f_{1}+f_{2}=1$). It will also
prove useful to introduce the \emph{conjugate} lattice by the
following definition: two conjugated binary lattices are related to
one another through the replacement $J=1,2\rightleftarrows 2,1$
(that is, $\mu _{1},~f_{1}\rightleftarrows \mu _{2},~f_{2}$)
interchanging the materials along with their filling fractions. The
conjugated lattices are referred to below as 1/2 and 2/1 lattices,
with the matrix material put first. Note that the exact effective
speeds in conjugated lattices are in general certainly different,
$c_{\left( 1/2\right) }\neq c_{\left( 2/1\right) }$, except for
particular symmetric lattice configurations, see \S \ref{sec5.1.2}.

The PWE estimate (\ref{21}$_{2}$) of the effective speed $c$ reduces to the
form%
\begin{equation}
c_{\mathrm{PWE}}^{2}=\frac{1}{\left\langle \rho \right\rangle }\left( \mu
_{1}f_{1}+\mu _{2}f_{2}-\frac{f_{1}f_{2}\left( \mu _{1}-\mu _{2}\right) ^{2}%
}{\mu _{1}+\mu _{2}}\right) ,  \label{24}
\end{equation}%
which, by definition, depends only on the filling fractions $f_{J}$ and is
not sensitive to the inclusion shape. It is also evident that (\ref{24}) is
invariant under the interchange $J=1,2\rightleftarrows 2,1$, i.e., $c_{%
\mathrm{PWE}}$ is the same for the two {conjugated} binary lattices.

The MM estimate $c_{\mathrm{MM}}^{2}$ for the two-phase square lattice is
given by (\ref{M18.1}) with $J=1,2$. It is not invariant to interchanging $%
J=1,2\rightleftarrows 2,1,$ i.e. the effective speed for each of the
conjugated lattices has its own MM estimate $c_{\left( 1/2\right) }\approx
c_{\mathrm{MM}}^{\left( 1/2\right) }$ and $c_{\left( 2/1\right) }\approx c_{%
\mathrm{MM}}^{\left( 2/1\right) }$ (where $c_{\mathrm{MM}}^{\left(
1/2\right) }=c_{\mathrm{MM}}^{\left( 2/1\right) }$ for the symmetric
configurations).

The estimate obtained by means of the multiple-scattering theory (MST) \cite%
{MLWS,TS,SMLW,TS1} is, for the 1/2 lattice,%
\begin{gather}
c_{\left( 1/2\right) }^{2}\approx \left[ c_{\mathrm{MST}}^{\left( 1/2\right)
}\right] ^{2}=\frac{\mu _{1}}{\left\langle \rho \right\rangle }\left( \frac{%
\mu _{1}+\mu _{2}-\left( \mu _{1}-\mu _{2}\right) f_{2}}{\mu _{1}+\mu
_{2}+\left( \mu _{1}-\mu _{2}\right) f_{2}}\right) ,  \notag \\
J=1\ \mathrm{is}\ \mathrm{matrix,\ }J=2\ \mathrm{is}\ \mathrm{inclusion.}
\label{25}
\end{gather}
Interchanging the indices $J=1,2\rightleftarrows 2,1$ in (\ref{25}) provides
the MST estimate for the conjugated 2/1 lattice as%
\begin{eqnarray}
c_{\left( 2/1\right) }^{2} &\approx &\left[ c_{\mathrm{MST}}^{\left(
2/1\right) }\right] ^{2}=\frac{\mu _{2}}{\left\langle \rho \right\rangle }%
\left( \frac{2\mu _{1}-\left( \mu _{1}-\mu _{2}\right) f_{2}}{2\mu
_{2}+\left( \mu _{1}-\mu _{2}\right) f_{2}}\right) ,  \notag \\
J &=&2\ \mathrm{is}\ \mathrm{matrix,\ }J=1\ \mathrm{is}\ \mathrm{inclusion.}
\label{26}
\end{eqnarray}%
The MST estimate defines distinct values of $c_{\mathrm{MST}}$ for the two
conjugated lattices. The choice as to which of the MST formulas (\ref{25}), (%
\ref{26}) to apply to a given binary lattice depends crucially on the
designation of the two constituent materials as 'matrix' and 'inclusion'.
There is no ambiguity for simple configurations where one of the materials
('matrix') circumvents the unit-cell boundary and the other is enclosed
within ('inclusion'). However, in the case of a symmetric lattice
configuration, for which two conjugated lattices are equivalent, Eqs. (\ref%
{25}) and (\ref{26}) provide two starkly different MST
approximations of a single exact value $c_{\left( 1/2\right)
}=c_{\left( 2/1\right) },$ see further \S \ref{sec5.1.2}.

Note that the explicit expressions (\ref{25}), (\ref{26}) actually have a
long record in micromechanics, see \cite{BM,MMP,PA}. In particular, they are
the Hashin-Shtrikman bounds (respectively, upper and lower at $\mu _{1}>\mu
_{2}$ or vice versa at $\mu _{1}<\mu _{2}$) obtained by the variational
approach for a binary composite of a matrix material $J=1$ or $2$ with
statistically homogeneous 
inclusions of material $J=2$ or $1$, see \cite{H}. The relation of these
bounds to periodic structures may not be generally obvious. At the same
time, for the two-phase lattices, it is easy to verify explicitly that the
PWE estimate (\ref{24}) is always enclosed between (\ref{25}) and (\ref{26}%
), and that the upper Hashin-Shtrikman bound is never greater than the PWE
bound (\ref{21}$_{1}$) for the two-phase case; however, the same is not
always true for the MM estimate (\ref{M18.1}) with $J=1,2$. One more general
result from the theory of 2D two-phase composites is noteworthy, which is
Keller's duality relation \cite{KF,M} for the effective shear coefficients $%
\mu _{\mathrm{eff}}$ of two $\emph{reciprocal}$ lattices $\left( \mu
_{1},\mu _{2}\right) $ and $\left( \mu _{2},\mu _{1}\right) $ obtained from
one another by interchanging $\mu _{1}\rightleftarrows \mu _{2}$ while
keeping the concentrations $f_{1,2}$ intact (cf. the definition of
conjugated lattices). For the isotropic case in hand, this relation yields
the identity%
\begin{equation}
\left\langle \rho \right\rangle ^{2}c_{\left( \mu _{1},\mu _{2}\right)
}c_{\left( \mu _{2},\mu _{1}\right) }=\mu _{1}\mu _{2}.  \label{27}
\end{equation}%
Among the above-mentioned estimates of $c$, the MST formulas (\ref{25}), (%
\ref{26}) satisfy (\ref{27}), while the PWE and MM approximations (\ref{24})
and (\ref{M18.1}) do not. Note that the $\widetilde{\mathrm{MM}}$ estimate (%
\ref{M20}) does satisfy (\ref{27}); however, the numerical tests (omitted
from the graphical data below to avoid its overloading) show that fitting of
$c$ by the MM estimate $c_{\mathrm{MM}}$ given by (\ref{M18.1}) is always
better than by $c_{\widetilde{\mathrm{MM}}}\left( \leq c_{\mathrm{MM}%
}\right) $~given by (\ref{M20}). The degree to which it is better is often
quantitative small, but then the departure of $c_{\mathrm{MM}}$ from the
duality identity (\ref{27}) is equally small. A greater accuracy of (\ref%
{M18.1}) than of (\ref{M20}) extends to the case of $J>2,$ where (\ref{M20})
has no methodological advantage of satisfying (\ref{27}) since the latter is
no longer relevant. Thus, all in all the MM estimate in the form (\ref{M18.1}%
) appears to be preferable to (\ref{M20}).

\subsubsection{Examples}

\label{sec5.1.2}

\begin{figure}[h] \centering
\includegraphics{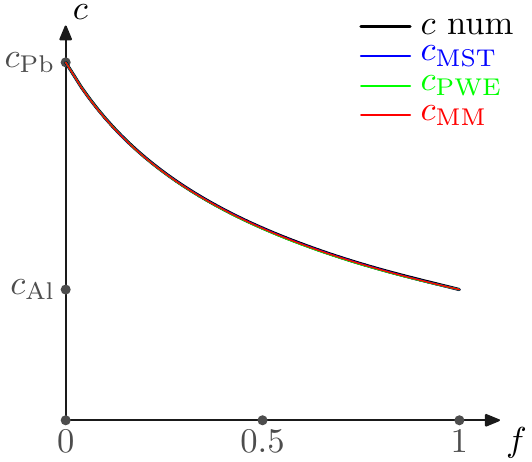}
{\caption{Effective speed $c$ versus concentration $f_{\mathrm{Al}}$
for conjugated Al/Pb and Pb/Al lattices of 0${{}^\circ}$-oriented
rods. The numerical curves for both lattices (computed via (\ref
{17.2}) with $N=25$ and $m=10$), the PWE estimate (\ref{24}), the MM
estimate (\ref{M18.1}) and the MST approximations (\ref{25}),
(\ref{26}) all merge at the scale of the plot.}\label{fig1}}
\end{figure}

In this subsection, a comparison between the numerical evaluation of the
effective speed $c$ and its different estimates is demonstrated for several
examples of a square lattice of parallel square rods embedded in a matrix
and oriented at an angle of 0$%
{{}^\circ}%
$ or 45$%
{{}^\circ}%
$ to the translation vectors. Such configurations of phononic crystals have
been studied, e.g., in \cite{GV,WLL,F-M,F-MM,M-B}. It is clear that the MST
estimate of \cite{MLWS,TS,SMLW,TS1}, though derived for cylindrical
inclusions, should be equally viable for square ones since it describes the
quasistatic limit. If the contrast of matrix and inclusion shear
coefficients is relatively low, then so is the difference between the two
values of the effective speed $c$ for the two conjugated lattices. In this
case, the PWE, MM and MST estimates (\ref{24}), (\ref{M18.1}) and (\ref{25}%
)-(\ref{26}) all yield close values that provide a good approximation of $c$
in either of the conjugated configurations. This is exemplified in Fig. 1
for Al and Pb phases with the material constants $\rho _{\mathrm{Al}}=2.7$, $%
\rho _{\mathrm{Pb}}=11.6~$g/cm$^{3}$ and$\ \mu _{\mathrm{Al}}=26$, $\mu _{%
\mathrm{Pb}}=14.9$ GPa. Note that the series (\ref{17.2}) needs only about $%
j\sim 7$ modes ($N\sim 15$) and $m\sim 5$ terms for accurate calculation of
the numerical curve $c\left( f_{\mathrm{Al}}\right) $ (the larger values of $%
N$ and $m$ indicated in the caption were taken for better precision).

\begin{figure}[h]
\begin{minipage}[h]{0.99\linewidth}
\centering\includegraphics{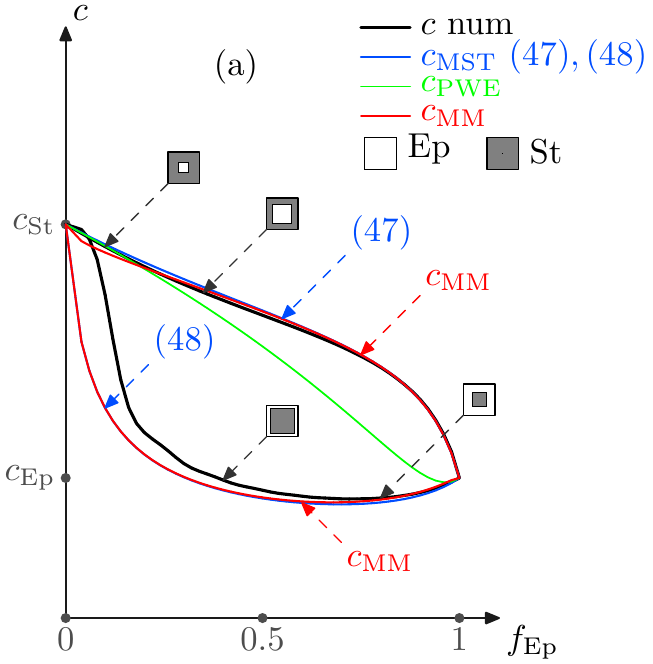}  \\
\end{minipage}
\vfill
\begin{minipage}[h]{0.99\linewidth}
\centering\includegraphics{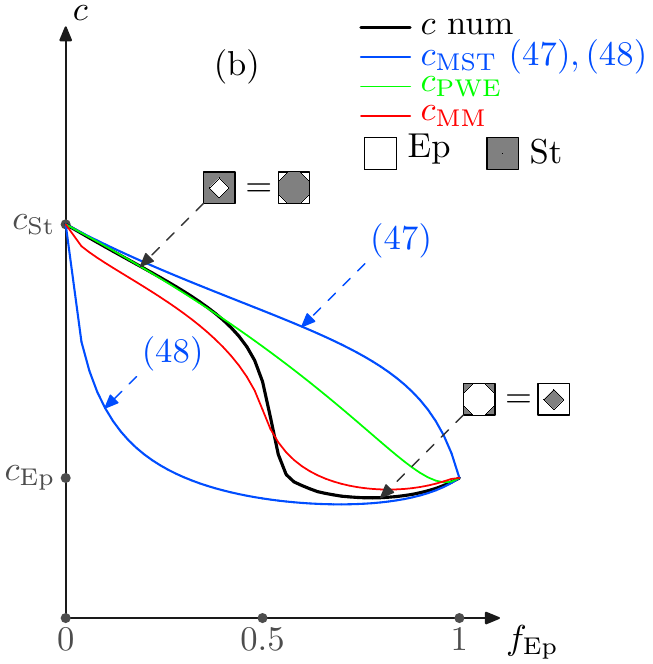}  \\
\end{minipage}
\caption{Effective speed (a) for the conjugated St/Ep and Ep/St
lattices of 0${{}^\circ}$-oriented rods and (b) for the symmetric
St/Ep lattice of 45${{}^\circ}$-rotated rods. Numerical curves
$c\left( f_{\mathrm{Ep}}\right) $ are computed via (\ref{17.2}) with
$N=841$ and $m=150$; the PWE estimate $c_{ \mathrm{PWE}}$ is given
by (\ref{24}); the MM estimate $c_{ \mathrm{MM}}$ is given by
(\ref{M18.1}); the MST approximations
$c_{\mathrm{MST}}^{(\mathrm{St/Ep})}$ and
$c_{\mathrm{MST}}^{(\mathrm{Ep/St})}$ are given by (\ref{25}) and
(\ref{26}) with $J=1=\mathrm{St}$, $J=2=\mathrm{Ep.}$} \label{fig2}
\end{figure}

Addressing the high-contrast case, consider two examples of binary materials
with a 'medium' and 'drastic' contrast: one consisting of steel ($\equiv $
St) and epoxy ($\equiv $ Ep), and the other of steel and rubber ($\equiv $
R). Their material constants are $\rho _{\mathrm{St}}=7.8$,\ $\rho _{\mathrm{%
Ep}}=1.14$, $\rho _{\mathrm{R}}=1.14\ $g/cm$^{3}\ $and $\mu _{\mathrm{St}%
}=80 $, $\mu _{\mathrm{Ep}}=1.48$, $\mu _{\mathrm{R}}=4\cdot 10^{-5}\ $GPa.
The results for the St/Ep and Ep/St conjugated lattices of 0$%
{{}^\circ}%
$-oriented rods are shown in Fig. 2a, and the results for the St/R and R/St
lattices are shown in Fig. 3a. It is seen that the two numerical curves $%
c\left( f\right) ,$ plotted for each conjugated pair as a function of
concentration of the same (say, softer) material, have quite different
trajectories between the fixed end points. The physical reason is obvious:
the effective speed $c$ is indeed strongly affected by a small concentration
of a highly contrasting component when this forms a 'network' breaking up
connectivity of the volume-dominating component. On the numerical side,
given the 'medium-contrast' case of steel-epoxy composite, Eq. (\ref{17.2})
provides a reasonable approximation of $c\left( f\right) $ when taken with $%
j=7$ modes ($N\sim 15$) and $m\sim 50$ terms (compare with the above Al-Pb
case). About this number of modes and terms in Eq. (\ref{17.2}) is also
sufficient to capture the shape of the curve $c\left( f\right) $ for the
'drastic-contrast' steel-rubber structure but only if rubber is an inclusion
located inside the cell. Markedly more numerical effort is required when
rubber is the matrix material distributed along the unit-cell boundaries -
in this case no less than $j=12$ modes ($N\sim 25$) and $m\sim 150$ terms in
Eq. (\ref{17.2}) are needed to obtain good accuracy (see \S\ 4). Note that
formally reducing $\mu _{\mathrm{R}}$ to zero causes no discernible changes
at the scale of Figs. 3, 4.

\begin{figure}[h]
\begin{minipage}[h]{0.99\linewidth}
\centering\includegraphics{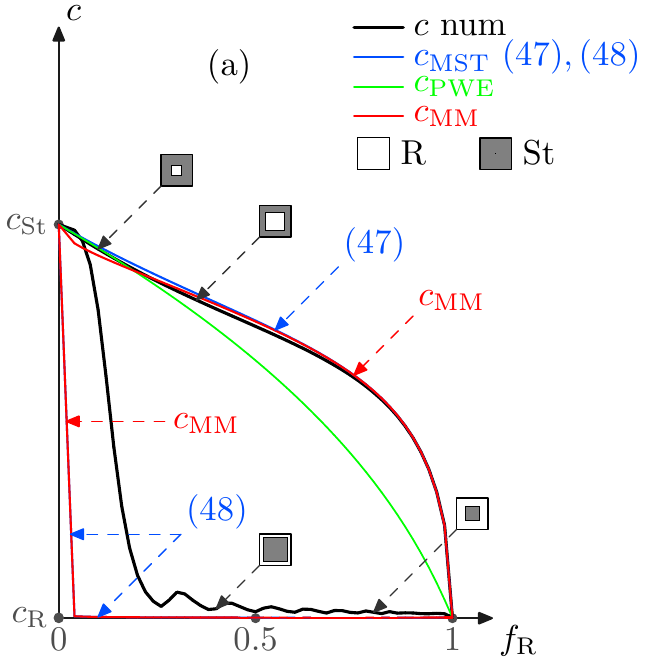}  \\
\end{minipage}
\vfill
\begin{minipage}[h]{0.99\linewidth}
\centering\includegraphics{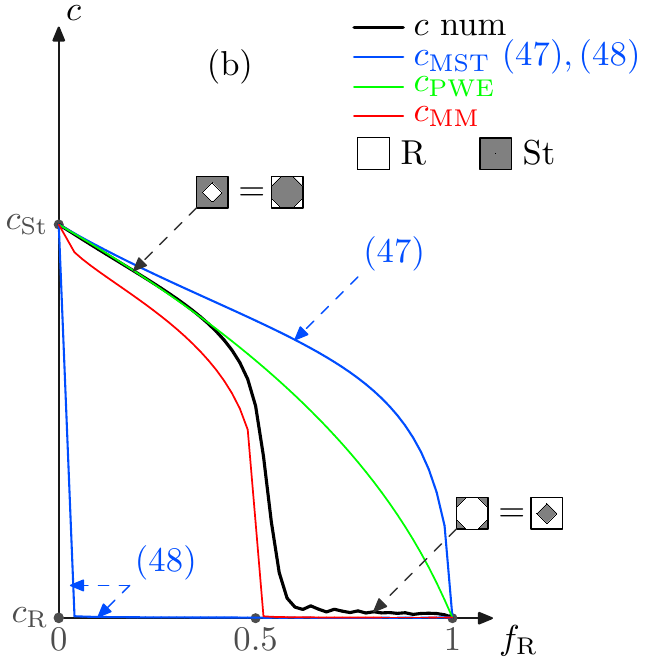}  \\
\end{minipage}
\caption{The same as in Fig. \ref{fig2} (a,b) but for St/R and R/St
lattices. Numerical
curves $c\left( f_{\mathrm{R}}\right) $ are computed via (\ref{17.2}) with $%
N=29$ and $m=150$. Note that $c_{\mathrm{R}}$ is not distinguishable
from $0$ at the scale of the vertical axis.} \label{fig3}
\end{figure}

Let us now examine the PWE, MM and MST estimates of $c$ for the above
examples. It is evident that a single curve of the PWE estimate, which
depends only on volume fraction and disregards geometrical details (see \S %
\ref{sec5.1.1}), cannot fit two markedly different curves of
conjugated lattices. As noted in \S \ref{sec3}, it must be more
accurate when the stiff component is volumetrically dominant over
the soft one rather than when the situation is reversed. This is
what is observed in Figs.\ 2a and 3a. It is also seen that the MM
and MST estimates provide a fairly close evaluation of $c,$ which
fits very well the whole numerical curve of $c$ for St/Ep and St/R
lattices (soft rods in stiff matrix); however, they lose accuracy
for the conjugated, Ep/St and R/St lattices (stiff rods in soft
matrix), specifically when the rod concentration $f_{\mathrm{St}}$
($=1-f_{\mathrm{Ep,R}}$) is close to 1. Regarding MST, this is in
agreement with the remark made on its derivation in
\cite{MLWS,TS,SMLW,TS1} that the MST estimate does not fully account
for the multiple interactions and hence may be error prone in the
case of densely packed stiff inclusions. Thus, in the latter case,
the PWE estimate is preferable to two others, as illustrated in
Fig.\ 2a and especially in Fig. 3a.

\begin{figure}[h]
\begin{minipage}[h]{0.99\linewidth}
\centering\includegraphics{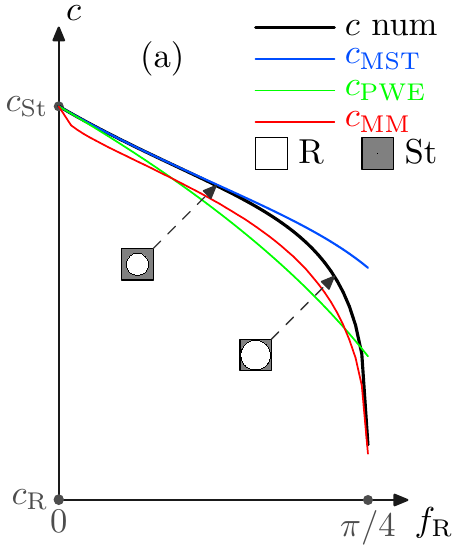}  \\
\end{minipage}
\vfill
\begin{minipage}[h]{0.99\linewidth}
\centering\includegraphics{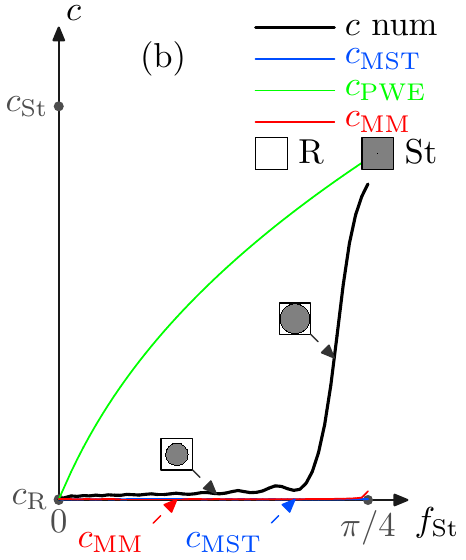}  \\
\end{minipage}
\caption{Effective speed as a function of concentration of
inclusions in (a) St/R and (b) R/St conjugated lattices of circular
cylinders in a matrix.
Numerical curves $c\left( f_{\mathrm{R}}\right) $ and $c\left( f_{\mathrm{St}%
}\right) $ are computed via (\ref{17.2}) with $N=29$ and $m=150$.}
\label{fig4}
\end{figure}

Consider next similar structures but with 45$%
{{}^\circ}%
$-rotated rods, which is the case where the two conjugated lattices coincide
into one symmetric configuration. The corresponding dependence of the
effective speed versus concentration $c\left( f\right) $ has a single-valued
approximation for each of the PWE and MM estimates, whereas the MST estimate
still defines two different approximations (\ref{25}) and (\ref{26}) for the
single curve $c\left( f\right) $. Comparing these estimates displayed
alongside the numerical curve $c\left( f\right) $ in Figs. 2b and 3b shows
that the PWE estimate is the most accurate so long as the stiff component is
volume-dominant; the MM estimate provides the best 'overall' fit; and each
of the MST approximations works over less than a half of the range while
mismatching markedly the other half.

Finally, we consider the case of cylindrical inclusions. Results for the
steel - rubber conjugate lattices with circular rods are presented in Fig.
4. It is instructive to observe the similarity of the dependences $c\left(
f\right) $ on the concentration of inclusions $f=f_{\mathrm{St}}$ and $f_{%
\mathrm{R}},$ which are displayed in Figs. 4a and 4b, to the two
corresponding 'halves' of the corresponding curves for square rods in Fig.
3b.

\subsection{Three-phase lattices}

\label{sec5.2}

\subsubsection{Estimates}

\label{sec5.2.1}

Consider a 2D square lattice similar to above but with a coated inclusion.
Such nested structures have received much attention lately in relation to
modelling locally resonant phononic crystals, e.g. \cite{LCS, LPDV}. The PWE
and MM estimates of the effective speed $c$ for this case are given by Eqs. (%
\ref{21}$_{2}$) and (\ref{M18.1}) with $\left\langle \cdot \right\rangle
=\sum_{J}\left( \cdot \right) _{J}f_{J}$ and $J=1,2,3$. If the concentration
$f_{J}$ of one of the constituent materials tends to zero, the MM estimate (%
\ref{M18.1}) for three constituents certainly tends to that for two
remaining constituents; whereas the PWE estimate (\ref{21}$_{2}$) with, say,
$f_{3}\rightarrow 0$ tends to its form for the pair $J=1,2$ only if the
'vanishing' material is neither the stiffest nor the softest one, i.e. if $%
\mu _{3}\neq \mu _{\min },$ $\mu _{\max }$.

As a MST counterpart, we adopt the generalization of (\ref{25}) that is
well-known in micromechanics as the Kuster-Toks\"{o}z formula (closely
related to Hashin-Shtrikman bounds) for 2D fluids with small concentration
of different inclusions \cite{KT, B}. More recently, it was used for a
periodic structure of different cylinders in a fluid matrix \cite{TS,TS1}.
The formula for the 2D configurations considered here is%
\begin{eqnarray}
c^{2} &\approx &c_{\mathrm{MST}}^{2}=\frac{\mu _{1}}{\left\langle \rho
\right\rangle }\left( \frac{1-\sum\nolimits_{J=2}^{3}f_{J}\frac{\mu _{1}-\mu
_{J}}{\mu _{1}+\mu _{J}}}{1+\sum\nolimits_{J=2}^{3}f_{J}\frac{\mu _{1}-\mu
_{J}}{\mu _{1}+\mu _{J}}}\right) ,  \label{28} \\
J &=&1\mathrm{\ is\ matrix,\ }J=2,3\mathrm{\ are\ inclusions.}  \notag
\end{eqnarray}
The MST estimate (\ref{28}) coincides with the binary formula (\ref{25}) if
any one of the inclusion concentrations $f_{2}$ or $f_{3}$ is zero. On the
other hand, (\ref{28}) does not tend to either of (\ref{25}) and (\ref{26})
as the matrix concentration $f_{1}$ tends to zero (which is not surprising
since the Kuster-Toks\"{o}z is not recommended at low matrix concentration
\cite{B2}).

\subsubsection{Examples}

\label{sec5.2.2}

\begin{figure}[h]
\begin{minipage}[h]{0.99\linewidth}
\centering\includegraphics{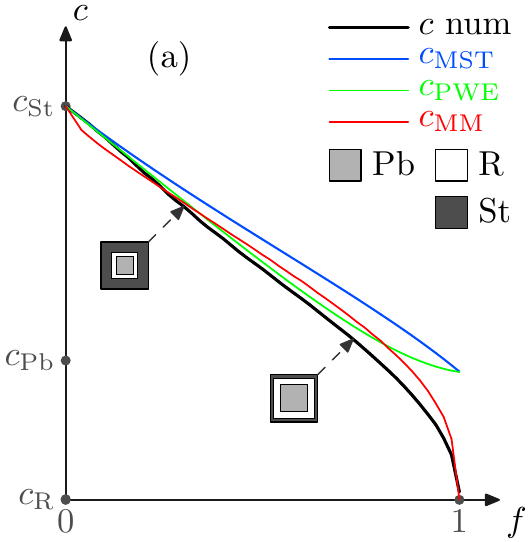}  \\
\end{minipage}
\vfill
\begin{minipage}[h]{0.99\linewidth}
\centering\includegraphics{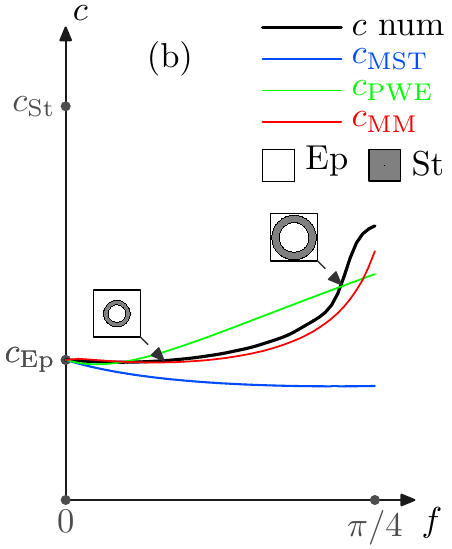}  \\
\end{minipage}
\caption{Effective speed $c\left( f\right) $ for three-phase
lattices where $ f $ is given by (\ref{29}) with $\alpha =4/9$: (a)
Pb/Ru/St structure of coated square rods and (b) Ep/St/Ep structure
of cylindrical annuli. Numerical curves are computed via
(\ref{17.2}) with $N=29$ and $m=150$} \label{fig5}
\end{figure}

Denote the filling fraction of a coated inclusion in a matrix ($J=1$) by $f$
and set the filling fractions of the skin ($J=2$) and core ($J=3$) materials
as%
\begin{equation}
f_{2}=\alpha f\ \ (\mathrm{skin}),\ f_{3}=\left( 1-\alpha \right) f\ \ (%
\mathrm{core})\ \Rightarrow \ f_{2}+f_{3}=f.  \label{29}
\end{equation}%
The effective speed $c$ of the three-phase composite is now a function of
the single variable $f=1-f_1$.

Motivated by \cite{LCS, LPDV}, we first examine the case of a soft coating
(skin) material. Consider the square St/R/Pb lattice of square lead ($\equiv
$Pb) rods coated by rubber ($\equiv $R) which are embedded in steel matrix
(Fig. 5a). The value of $c\left( f\right) $ at $f=0$ is obviously the speed
in the matrix, $c\left( 0\right) =c_{\mathrm{St}}$. The opposite limit value
of $c\left( f\right) $ at $f=1$ is equal to the effective speed $c_{\mathrm{%
R/Pb}}\left( f_{\mathrm{R}}\right) $ in the binary R/Pb lattice of lead rods
embedded in the rubber matrix with the volume fractions fixed by (\ref{29})
as $f_{\mathrm{R}}=\alpha $ and $f_{\mathrm{Pb}}=1-\alpha $. Once $f_{%
\mathrm{R}}$ is not too small, $c_{\mathrm{R/Pb}}\left( f_{\mathrm{R}%
}\right) $ should be close to $c_{\mathrm{R}}$ (see Fig. 3a), which
therefore implies that $c\left( f\right) $ in the St/R/Pb structure has a
very small value in the limit $f\rightarrow 1$. This is observed in Fig. 5a
(where $\alpha =4/9$). It is also seen that the PWE and MST estimates (\ref%
{21}$_{2}$) and (\ref{28}) of $c(f)$ do not describe this behaviour of $c(f)$
at $f\rightarrow 1$ and overestimate $c\left( 1\right) $ (by an incidentally
close value which is neither PWE nor MST estimate of $c_{\mathrm{R/Pb}%
}\left( f_{\mathrm{R}}\right) $, as pointed out in \S \ref{sec5.2.1}
above). By contrast, the MM estimate (\ref{M18.1}) provides a good
fit for the whole curve $c\left( f\right) $ including the critical
region $f\rightarrow 1$. This is because Eq. (\ref{M18.1}) captures
the 'insulating' effect of a small concentration of soft material
which drastically decreases the effective speed when this material
extends throughout the unit-cell boundary, see \S \ref{sec3.2}.

Another case of interest is when the matrix material coincides with that of
the rod core, which means that the rod coatings are simply spacers
separating the same material. Figure 5b demonstrates the dependence of the
effective speed $c$ on the concentration $f$ of stiff (steel) cylindrical
annuli embedded in a soft (epoxy$\equiv $Ep) material. The shape of the
curve $c\left( f\right) $ can be shown to change only slightly if the steel
spacers are square instead of circular. It is seen from Fig. 5b that the
basic outline of this curve is again best approximated by the MM estimate.

\section{Conclusion}

\label{sec6}

The paper uses the PWE approach and a newly developed MM approach, based on
the monodromy matrix, to derive the new estimates of the effective
shear-wave speed $c$ in 2D periodic lattices. The estimates are compared
with the known MST approximations and with the numerical data for a number
of examples of two- and three-phase square lattices. The main findings are
listed in the Introduction. The results for effective velocities of the
vector waves in the 3D lattices are to be reported elsewhere. It is worth
pointing out that the obtained PWE and MM estimates are also valid for the
gradient-index, or functionally graded, materials (for which the MST is
irrelevant). In conclusion, the combination of the perturbation theory with
the PWE and MM techniques, which is elaborated in this paper, is hoped to
lend an efficient tool for a broad range of problems concerned with periodic
composites, phononic crystals and metamaterials.

\noindent \textbf{Acknowledgement.} This work has been supported by the
grant ANR-08-BLAN-0101-01 from the Agence Nationale de la Recherche and by
the project SAMM from the cluster Advanced Materials in Aquitaine. A.N.N.
acknowledges the support by the Centre National de la Recherche
Scientifiqueis.\medskip

\noindent \textbf{APPENDIX. Convergence of (\ref{17.1}}$_{2}$\textbf{): a
strict example }

\underline{\textbf{Sufficient condition on }$\mu (\mathbf{x})$\textbf{.}}%
\textbf{\ }Our objective is to provide a rigorous example of a class of
functions $\mu \left( \mathbf{x}\right) \equiv \mu _{0}+\mu _{\Delta }\left(
\mathbf{x}\right) $ that guarantee convergence for $M(\mathbf{\kappa })=\mu
_{0}^{-1}\sum\nolimits_{n=0}^{\infty }\left( \left( -\mathbf{C}\right) ^{n}%
\mathbf{f,f}\right) $, and thus validate application of this series for
computing the effective parameters $\mu _{\mathrm{eff}}(\mathbf{\kappa })$
and $c^{2}(\mathbf{\kappa })$. To do so, we begin by formulating a
sufficient condition on $\mu \left( \mathbf{x}\right) $ to fulfill the
sufficient condition $\left\Vert \mathbf{C}(\mu _{0})\right\Vert <1$ for
convergence of (\ref{17.1}$_{2}$) as $m\rightarrow \infty $. Note that the
matrix $\mathbf{\mathbf{C}}$ can be written as%
\begin{gather}
\mathbf{C}=\mu _{0}^{-1}\sum\nolimits_{\widetilde{\mathbf{g}}\in \Gamma }%
\widehat{\mu }_{\Delta }\left( \widetilde{\mathbf{g}}\right) \mathbf{J}_{%
\widetilde{\mathbf{g}}}\ \ \mathrm{where\ }\mathbf{J}_{\widetilde{\mathbf{g}}%
}\mathrm{:}  \notag \\
J_{\widetilde{\mathbf{g}}}\left[ \mathbf{\mathbf{g,g}}^{\prime }\right] =%
\begin{cases}
\frac{\mathbf{g}}{\left\vert \mathbf{g}\right\vert }\cdot \frac{\mathbf{g}%
^{\prime }}{\left\vert \mathbf{g}^{\prime }\right\vert } & \mathrm{if}%
\mathbf{\mathbf{\ \widetilde{\mathbf{g}}=\mathbf{g-g}^{\prime },}} \\
0\  & \mathrm{otherwise}\mathbf{.}%
\end{cases}
\label{A1}
\end{gather}%
It is seen from (\ref{A1}) that $\left\Vert \mathbf{\mathbf{C}}\right\Vert
\leq \mu _{0}^{-1}\sum\nolimits_{\widetilde{\mathbf{g}}\in \Gamma
}\left\vert \widehat{\mu }_{\Delta }\left( \widetilde{\mathbf{g}}\right)
\right\vert $ since $\left\Vert \mathbf{J}_{\widetilde{\mathbf{g}}%
}\right\Vert \leq 1,$ which in turn is because all its nonzero elements $J_{%
\widetilde{\mathbf{g}}}\left[ \mathbf{\mathbf{g,g}}^{\prime }\right] $
occupy a single particular diagonal and satisfy $\left\vert J_{\widetilde{%
\mathbf{g}}}\left[ \mathbf{\mathbf{g,g}}^{\prime }\right] \right\vert \leq 1$%
. Hence the sufficient convergence condition $\left\Vert \mathbf{C}%
\right\Vert <1$ may be eased to%
\begin{multline}
\left( \left\Vert \mathbf{C}(\mu _{0})\right\Vert \leq \right) ~\mu
_{0}^{-1}\sum\nolimits_{\widetilde{\mathbf{g}}\in \Gamma }\left\vert
\widehat{\mu }_{\Delta }\left( \widetilde{\mathbf{g}}\right) \right\vert
=1-\frac{\left\langle \mu \right\rangle }{\mu _{0}}   \\
+\frac{1}{\mu _{0}}\sum\nolimits_{\mathbf{g}\neq
\mathbf{0}}\left\vert \widehat{\mu }\left( \mathbf{g}\right)
\right\vert  \equiv \Theta _{\mu _{0}}<1\ \mathrm{for}\ \ \mu
_{0}\geq \left\langle \mu \right\rangle. \label{A2}
\end{multline}
In other words, for those $\mu (\mathbf{x})$ which satisfy
\begin{equation}
\sum\nolimits_{\mathbf{g}\neq \mathbf{0}}\left\vert \widehat{\mu }\left(
\mathbf{g}\right) \right\vert <\left\langle \mu \right\rangle   \label{A3}
\end{equation}%
there always exists a choice of $\mu _{0}\geq \left\langle \mu \right\rangle
$ which ensures $\left\Vert \mathbf{C}(\mu _{0})\right\Vert <1$ and hence
guarantees convergence of \textbf{(}\ref{17.1}$_{2}$\textbf{) }to $M\left(
\mathbf{\kappa }\right) $. The remainder of the series \textbf{(}\ref{17.1}$%
_{2}$\textbf{) }with $\mu _{0}\geq \left\langle \mu \right\rangle $ may be
estimated as follows%
\begin{gather}
\left\vert \mu _{0}^{-1}\sum\nolimits_{n=m+1}^{\infty }\left( \left( -%
\mathbf{C}\right) ^{n}\mathbf{f,f}\right) \right\vert \leq \frac{\left\Vert
\mathbf{f}\right\Vert ^{2}}{\mu _{0}}\sum\nolimits_{n=m+1}^{\infty
}\left\Vert \mathbf{C}\right\Vert ^{n}  \notag \\
<\frac{\left\langle \mu \right\rangle ^{2}}{\mu _{0}}\frac{\Theta _{\mu
_{0}}^{m+1}}{1-\Theta _{\mu _{0}}}=\frac{\left\langle \mu \right\rangle
^{2}\Theta _{\mu _{0}}^{m+1}}{\left\langle \mu \right\rangle -\sum\nolimits_{%
\mathbf{g}\neq \mathbf{0}}\left\vert \widehat{\mu }\left( \mathbf{g}\right)
\right\vert },  \label{A4}
\end{gather}
where it has been used that $\left\Vert \mathbf{C}\right\Vert \leq $ $\Theta
_{\mu _{0}}$ by (\ref{A2}) and that%
\begin{eqnarray}
\left\Vert \mathbf{f}\right\Vert  &=&\sqrt{\left\langle \left\vert f\left(
\mathbf{x}\right) \right\vert ^{2}\right\rangle }\leq \max\limits_{\mathbf{x}%
}\left\vert f\left( \mathbf{x}\right) \right\vert   \label{A5} \\
&\leq &\sum\nolimits_{\mathbf{g}\neq \mathbf{0}}\left\vert \widehat{\mu }%
\left( \mathbf{g}\right) e^{i\mathbf{g\cdot x}}\frac{\mathbf{g}}{\left\vert
\mathbf{g}\right\vert }\cdot \mathbf{\kappa }\right\vert <\left\langle \mu
\right\rangle   \notag
\end{eqnarray}
for$\ f\left( \mathbf{x}\right) =\sum\nolimits_{\mathbf{g}\neq \mathbf{0}}%
\widehat{f}\left( \mathbf{g}\right) e^{i\mathbf{g\cdot x}}$ by (\ref{12})
and (\ref{A3}). The least value of the residual sum (\ref{A4}) for all $\mu
_{0}\geq \left\langle \mu \right\rangle $ is achieved when $\Theta _{\mu
_{0}}$ is minimum, which is the case when $\mu _{0}=\left\langle \mu
\right\rangle $. Note that the average $\left\langle \mu \right\rangle $ of $%
\mu (\mathbf{x})$ satisfying (\ref{A3}) may well differ (be greater
or less) than the value $\overline{\mu }\equiv \frac{1}{2}\left( \mu
_{\max }+\mu _{\min }\right) ,$ which was argued in \S \ref{sec4} as
a numerically reliable choice of $\mu _{0}$ in
\textbf{(}\ref{17.1}$_{2}$\textbf{)}. There is indeed no
contradiction in this difference. First, recall that all the
conclusions of Appendix stem from only the sufficient conditions.
Second, as mentioned in \S \ref{sec4}, an advantage of taking
\textbf{(}\ref{17.1}$_{2}$\textbf{)} with $\mu _{0}=\overline{\mu }$
is that it yields the same formula (\ref{17.2}) for any profile $\mu
(\mathbf{x})$, but this choice of $\mu _{0}$ is not intended to
provide the fastest convergence for all possible profiles.

We still need to examine the restrictions on $\mu \left( \mathbf{x}\right) $
which are imposed by the derived sufficient condition (\ref{A3}). First of
all, by (\ref{A3}) $\mu (\mathbf{x})=\sum\nolimits_{\mathbf{g}}\widehat{\mu }%
\left( \mathbf{g}\right) e^{i\mathbf{g\cdot x}}\geq $ $\widehat{\mu }\left(
\mathbf{0}\right) -\sum\nolimits_{\mathbf{g}\neq \mathbf{0}}\left\vert
\widehat{\mu }\left( \mathbf{g}\right) \right\vert >0,$ i.e. only positive $%
\mu (\mathbf{x})$ are allowed as needed. Second, any $\mu (\mathbf{x})$
satisfying (\ref{A3}) must have a uniformly converging Fourier series and
hence be continuous. The latter is actually not a loss of generality in the
numerical context, even if we are mostly interested in the case of materials
with inclusions (i.e. with jumps of properties), because the calculations
deal with truncated Fourier series of $\mu (\mathbf{x})$ which in effect
replaces a possibly piecewise constant $\mu (\mathbf{x})$ by a continuous
profile. Thirdly, (\ref{A3}) implies that $\left\vert \mu (\mathbf{x}%
)-\left\langle \mu \right\rangle \right\vert \leq \left\langle \mu
\right\rangle $, i.e. $\mu (\mathbf{x})>0$ should not depart 'too far' from
its average $\left\langle \mu \right\rangle .$ When so, the matrix $\mathbf{%
I+C}$ is diagonal predominant and $\left\vert \mathbf{f}\right\vert $
decreases for large $\mathbf{g,}$ both furthering the truncation of the PWE
and of the power series in (\ref{17.1}$_{2}$). It is evident that the above
condition, which may be recast as $\mu _{\max }\leq 2\left\langle \mu
\right\rangle ,$ fits a fairly broad class of functions $\mu (\mathbf{x})$.

\underline{\textbf{Example.}} In constructing an explicit example of the
profile $\mu (\mathbf{x})$ which ensures convergence of \textbf{(}\ref{17.1}$%
_{2}$\textbf{)}, we consider one that emulates a high-contrast composite
with a small volume fraction of soft inclusions. For brevity of writing, let
$\mathbf{T}=\left[ -\pi ,\pi \right] ^{2}$ so that $\mathbf{x}=\left(
x_{1},x_{2}\right) ,$ $\mathbf{g}=\left( g_{1},g_{2}\right) $ with $x_{i}\in %
\left[ -\pi ,\pi \right] $ and $g_{i}=n_{i}$ ($n_{i}\in
\mathbb{Z}
$). Denote%
\begin{multline}
\varphi _{n_{1}n_{2}}\left( \mathbf{x}\right) =\psi _{n_{1}}\left(
x_{1}\right) \psi _{n_{2}}\left( x_{2}\right)
=
\\
\sum\nolimits_{\left\vert g_{1}\right\vert \leq n_{1};~\left\vert
g_{2}\right\vert \leq n_{2}}\widehat{ \varphi }_{n_{1}n_{2}}\left(
\mathbf{g}\right) e^{i\mathbf{g\cdot x}},
\\
\mathrm{where}\ \psi _{n}\left( x\right) \equiv \left( \cos x\right)
^{2n}=\sum\nolimits_{\left\vert g\right\vert \leq n}\widehat{\psi }
_{n}\left( g\right) e^{igx}.  \label{A6}
\end{multline}%
Since $\psi _{n}\left( \pi l\right) =1$ ($l\in
\mathbb{Z}
$) and $\psi _{n}\left( x\right) \rightarrow 0$ for $x\neq \pi l$ as $%
n\rightarrow \infty $, the function $\varphi _{n_{1}n_{2}}\left(
x_{1},x_{2}\right) $ for large $n_{1}$, $n_{2}$ tends to a 2D grid of narrow
unit peaks. Note also that $\widehat{\psi }_{n}\left( g\right) \geq 0\
\forall g$ and $\sum\nolimits_{\left\vert g\right\vert \leq n}\widehat{\psi }%
_{n}\left( g\right) =\psi _{n}\left( 0\right) =1$, whence $\widehat{\varphi }%
_{n_{1}n_{2}}\left( \mathbf{g}\right) \geq 0$ and $\sum\nolimits_{\left\vert
g_{1}\right\vert \leq n_{1},~\left\vert g_{2}\right\vert \leq n_{2}}\widehat{%
\varphi }_{n_{1}n_{2}}\left( \mathbf{g}\right) =1$. Using this $\varphi
_{n_{1}n_{2}}$, define the function $\mu (\mathbf{x})$ as follows:%
\begin{multline}
\mu (\mathbf{x})=\mu _{0}+\mu _{\Delta }\left( \mathbf{x}\right) :\\
\mu _{0}>A>0,\ \mu _{\Delta }\left( \mathbf{x}\right) =-A\varphi
_{n_{1}n_{2}}\left( \mathbf{x}\right) ,  \label{A7}
\end{multline}%
where $\mu _{0}$ and $A$ are some constants. From the above properties it
follows that%
\begin{multline}
\sum\nolimits_{\mathbf{g\neq 0}}\left\vert \widehat{\mu }\left( \mathbf{g}%
\right) \right\vert  =A\sum\nolimits_{\mathbf{g\neq 0}}\widehat{\varphi }%
_{n_{1}n_{2}}\left( \mathbf{g}\right) =
\\
A\left( 1-\widehat{\varphi }%
_{n_{1}n_{2}}\left( \mathbf{0}\right) \right)   \label{A8} <\mu
_{0}-A\widehat{\varphi }_{n_{1}n_{2}}\left( \mathbf{0}\right)
=\left\langle \mu \right\rangle .  \notag
\end{multline}%
Thus the function (\ref{A7}) satisfies the condition (\ref{A3}) sufficient
for convergence of \textbf{(}\ref{17.1}$_{2}$\textbf{)}.


\end{document}